\documentclass[11pt]{article}
\usepackage{jheppub} %
\usepackage{lineno}
\usepackage[dvipsnames]{xcolor} 
\usepackage{dirtytalk}
\usepackage{ytableau}
\usepackage[dvipsnames]{xcolor}
\usepackage{dsfont}
\usepackage{mathdots}
\usepackage{colortbl}
\usepackage{braket}
\usepackage{amsthm}
\usepackage{amsmath}
\usepackage{float}
\usepackage{bbm, dsfont}
\usepackage[normalem]{ulem}
\usepackage{slashed}
\usepackage{mathtools}
\usepackage{multirow}
\usepackage{cancel}
\usepackage{cleveref}
\usepackage{orcidlink}
\usepackage{enumerate}
\usepackage{fontawesome}
\usepackage{adjustbox}
\usepackage{relsize}
\usepackage{csquotes}
\usepackage{enumitem}
\usepackage{graphicx} %
\usepackage{amssymb}
\usepackage{amsmath}
\usepackage{amsthm}
\usepackage{xcolor}
\usepackage{algorithm}
\usepackage{algpseudocode}
\usepackage{hyperref}
\usepackage{cleveref}
\usepackage{booktabs}
\usepackage{multirow}
\usepackage{mathrsfs}

\DeclareFontFamily{U}{mathx}{\hyphenchar\font45}
\DeclareFontShape{U}{mathx}{m}{n}{
      <5> <6> <7> <8> <9> <10>
      <10.95> <12> <14.4> <17.28> <20.74> <24.88>
      mathx10
      }{}
\DeclareSymbolFont{mathx}{U}{mathx}{m}{n}
\DeclareMathSymbol{\bigtimes}{1}{mathx}{"91}

\usepackage[most]{tcolorbox}
\definecolor{block-gray}{gray}{0.85}
\newtcolorbox{thmblock}{colback=block-gray,grow to right by=5mm,grow to left by=5mm,
boxrule=0pt,boxsep=0pt,breakable}

\usepackage{tikz}
    \usetikzlibrary{positioning}
    \usetikzlibrary{arrows}
    \usetikzlibrary{shapes}
    \usepackage{pgfplots}
    \usetikzlibrary{calc}
    \usetikzlibrary{decorations.markings}
     \usetikzlibrary{decorations.pathmorphing}
    \tikzset{snake it/.style={decorate, decoration=snake}}
    \usetikzlibrary{matrix}

    \pgfplotsset{compat=1.18}

\definecolor{ourblue}{RGB}{0, 57, 120} %
\colorlet{ourlightblue}{ourblue!50!white}
\newcommand{\bs}[1]{\boldsymbol{#1}}
\newcommand{\eps}{\varepsilon}
\newcommand{\rd}{\mathrm{d}}

\def\beq{\begin{equation}}
\def\eeq{\end{equation}}
\def\bsp#1\esp{\begin{split}#1\end{split}}

\usetikzlibrary{shapes.geometric}

\newcommand{\smallpentagon}{%
\scalebox{0.3}{
\begin{tikzpicture}[scale=3.5] 
  \node[regular polygon, regular polygon sides=5, minimum size=0.9 cm, draw] {};
\end{tikzpicture}}
  }%

\def\beq{\begin{equation}}
\def\eeq{\end{equation}}
\def\bsp#1\esp{\begin{split}#1\end{split}}

\newtheorem{theorem}{Theorem}

\newtheorem{lemma}{Lemma}
\newtheorem{corollary}{Corollary}
\theoremstyle{definition}

\newtheorem{remark}{Remark}

\preprint{BONN-TH-2026-20}

\title{Numerical evaluation of one-loop integrals using quadrature rules}

\abstract{We introduce a numerical quadrature algorithm with stretched exponential convergence to evaluate one-loop integrals in dimensional regularisation in the causal Loop Tree Dual (cLTD) representation. The key feature that makes this possible is the fact that for Euclidean kinematics all singularities of the cLTD  integrand lie on a compact set in loop momentum space. We prove that our algorithm admits a stretched exponential convergence with the number of integrand evaluations, and we illustrate it by evaluating triangle, box and pentagon integrals at selected Euclidean kinematics points. We benchmark our implementation against analytic results as well as results obtained via Monte Carlo integration. We also provide a proof of concept that our algorithm can be extended to physical kinematics after subtraction of threshold singularities. 
}
\author[a%
]{Jürgen Dölz}
\emailAdd{doelz@ins.uni-bonn.de}
\affiliation[a]{Institute for Numerical Simulation, Universität Bonn, D-53115, Germany
}
\author[b]{Claude Duhr,}
\emailAdd{cduhr@uni-bonn.de}
\affiliation[b]{Bethe Center for Theoretical Physics and Cluster of Excellence ``Color meets Flavor'', Universit\"{a}t Bonn, D-53115 Bonn, Germany
}

\author[a%
]{Bartul Kovačić}
\emailAdd{kovacic@ins.uni-bonn.de}

\author[b]{Cathrin Semper,}
\emailAdd{csemper@uni-bonn.de}

\begin{document}

\maketitle

\newpage
\section{Introduction}
\label{sec.intro}

The evaluation of Feynman integrals is one of the cornerstones of modern approaches to making predictions for collider and gravitational wave experiments. Over the last decade, a lot of progress has been made on how to evaluate Feynman integrals (cf.,~e.g.,~ref.~\cite{Caola:2022ayt,Bourjaily:2022bwx} for recent reviews). Most of the effort has focused on multi-loop integrals, and the evaluation of one-loop integrals is often considered a solved problem. This, however, is not the case when working in dimensional regularisation~\cite{tHooft:1972tcz,Cicuta:1972jf,Bollini:1972ui}. It is indeed true that next-to-leading order corrections to physical observables only require results for integrals with at most four propagators up to terms that vanish with the dimensional regulator $\varepsilon$~\cite{Melrose:1965kb}. The relevant integrals are all known analytically, cf.~refs.~\cite{tHooft:1972tcz,Ellis:2007qk,Carrazza:2016gav,Barik:2026twt}. Higher-order corrections, however, involve products of lower-loop integrals, or between loop integrals and ultraviolet and/or infrared counterterms. A consistent treatment of dimensional regularisation then demands that also higher orders in $\varepsilon$ of one-loop integrals and one-loop integrals with five propagators are taken into account. There is an on-going effort to evaluate the relevant integrals analytically~\cite{DelDuca:2009ac,Dixon:2011ng,DelDuca:2011ne,DelDuca:2011jm,DelDuca:2011wh,Chavez:2012kn,Kozlov:2015kol,Kozlov:2016vqy,Bourjaily:2019exo,Syrrakos:2021nij,Henn:2022ydo,Haug:2022hkr,Haug:2023eqg,Ren:2023tuj,Duhr:2025azh} or semi-analytically~\cite{Badger:2022mrb,Buccioni:2023okz,Becchetti:2025osw,Bera:2025upg}, and it is known that all one-loop integrals can be evaluated in terms of polylogarithmic functions~\cite{Mork:2026wie}. The resulting analytic expressions, however, are often extremely complicated. This can pose an obstacle to having fast and stable numerical codes. A fully numerical treatment of these integrals may therefore be preferable.

Numerical approaches to evaluate Feynman integrals have a long history, and can be broadly classified into two different categories.\footnote{In addition there are various public codes that solve differential equations for Feynman integrals numerically or to derive series representations, cf.,~e.g.,~refs.~\cite{Liu:2022chg,Hidding:2020ytt,Armadillo:2022ugh,Prisco:2025wqs}.} The first consists in integrating the Feynman parameter representation and to use sector decomposition~\cite{Hepp:1966eg,Roth:1996pd,Binoth:2000ps,Bogner:2007cr,Heinrich:2008si} and contour deformation~\cite{Soper:1999xk,Nagy:2003qn,Nagy:2006xy,Gong:2008ww,Anastasiou:2007qb,Becker:2011vg,Becker:2012bi,Becker:2012bi,Becker:2012aqa,Buchta:2015wna,Capatti:2019edf} to extract the poles in $\varepsilon$ and to obtain the correct imaginary parts above threshold. There are public implementations of these algorithms into {pySECDEC} \cite{Borowka:2017idc} and {FIESTA} \cite{Smirnov:2021rhf}. The second consists in directly performing the integration in loop-momentum space. The integrals are rendered finite by constructing suitable ultraviolet (UV) and infrared (IR) counterterms~\cite{Becker:2012bi,Becker:2012aqa,Anastasiou:2018rib,Anastasiou:2020sdt,Anastasiou:2022eym,Anastasiou:2024xvk,Anastasiou:2025cvy,Haindl:2025jte,Anastasiou:2026jta}. Alternatively, one may construct a basis of integrals that are UV and IR finite~\cite{Panzer:2014gra,vonManteuffel:2014qoa,Gambuti:2023eqh,delaCruz:2024xsm,DeAngelis:2025agn,Dhani:2026cxx}. The correct imaginary parts are obtained by deforming the integration contour~\cite{Soper:1999xk,Binoth:2005ff,Nagy:2006xy}, or by subtracting the threshold singularities to render the integrals convergent~\cite{Kermanschah:2021wbk, Kermanschah:2025wlo, Kermanschah:2024utt,Kilian:2009wy,Aguilera-Verdugo:2020set}.

In the loop momentum representation, one naively needs to perform a fourfold integral per loop, i.e., at $L$ loops there is naively a $4L$-dimensional integral to perform. It is possible to reduce the dimensionality of the integral to $3L$. The basic idea goes back to Feynman's tree theorem~\cite{Feynman:1963qx,Feynman:1972fi}, which states that one-loop integrals can be written as a sum of cuts, loosely defined by taking residues at the energy poles of the propagators. Extensions of this idea to higher loops were first proposed by Soper in ref.~\cite{Soper:1999xk}, and culminated in the development of the \textit{Loop-Tree-Duality} (LTD) in ref.~\cite{Catani:2008xa}.
The derivations of both LTD and Feynman's tree theorem are very similar, but, in contrast to Feynman's tree theorem, the LTD integrand only involves single cuts. Since then the LTD framework has been developed further~\cite{Bierenbaum:2010cy,Runkel:2019yrs,Capatti:2019ypt,Bierenbaum:2012th}
  and successfully applied to one- and multi-loop numerical computations, including explicit contour-deformation and subtraction for integrals with physical threshold singularities~\cite{Catani:2008xa,Buchta:2015wna,Capatti:2019ypt,Capatti:2019edf}. The singularity structure of the corresponding integrand, including cancellation of spurious singularities, was studied in refs.~\cite{Buchta:2014dfa,Aguilera-Verdugo:2019kbz}, also from a perspective related to cosmological polytopes~\cite{Rios-Sanchez:2026qvi}.
  The cancellations can also be made manifest at the integrand level~\cite{Capatti:2020ytd,Capatti:2022mly,Aguilera-Verdugo:2020kzc, JesusAguilera-Verdugo:2020fsn,Bobadilla:2021pvr}. The resulting variant of LTD is sometimes referred to as the \emph{causal} LTD (cLTD) representation.
  More recently, this formalism has been applied to the numerical computation of physical scattering processes within Monte Carlo frameworks, notably through the Local Unitarity Approach~\cite{Capatti:2022tit, AbdulHameed:2026utd, Kermanschah:2024utt,Kermanschah:2025wlo}, which combines cLTD representations with local real-virtual cancellations, thereby implementing the Kinoshita-Lee-Nauenberg (KLN)~\cite{Kinoshita:1962ur,Lee:1964is} theorem at the integrand level of the differential cross section. 
  
The focus of this paper is the numerical evaluation of one-loop integrals in dimensional regularisation. We work with the LTD representation of one-loop integrals, and we may pick a finite basis by introducing suitable numerators for integrals with at least three propagators (tadpole and bubble integrals are known analytically to all orders in the dimensional regulator). We obtain in this way a three-dimensional convergent cLTD representation for these integrals which may be evaluated numerically (higher orders in $\eps$ can be implemented by adding a single integration over the propagator masses, cf.,~e.g., ref.~\cite{Duhr:2025azh}). Typically, in high-energy physics the method of choice to evaluate multi-dimensional integrals numerically is Monte Carlo (MC) integration, for example via the CUBA library~\cite{Hahn:2004fe}. MC integration, however, suffers from relative slow convergence with the number of integrand evaluations.
Our main result is that cLTD representations of one-loop integrals are ideally suited to apply modern quadrature methods. A naive application of Gauss-Legendre quadrature rules fails, because integrands of loop integrals necessarily have singularities. One may then approximate the integrals by building a mesh of hypercubes that gets refined as we approach the singular loci, and we apply Gauss-Legendre quadrature on each hypercube. At the same time, one may vary the order of the quadrature for each cube. Here we propose to use $hp$-quadrature~\cite{BABUSKA19905}, an effective way to refine the mesh of hypercubes with a varying quadrature order. To apply $hp$-quadrature, a certain set of conditions need to be met, and we show that one-loop integrals in Euclidean kinematics always satisfy these requirements. As a consequence, we can prove that our algorithm has a (stretched) exponential convergence.

Our paper is structured as follows: 
In section~\ref{sec.LTD}, we review the LTD representation for one-loop Feynman integrals, in particular how we can obtain at most four-dimensional convergent integral representations for integrals in dimensional regularisation. In section~\ref{sec.hp}, we introduce the $hp$-quadrature algorithm for numerical integration, and we show that the LTD representation of one-loop integrals is ideally suited to allow for efficient numerical integration via the $hp$-quadrature method. We also proof that our algorithm admits a stretched-exponential convergence with the number of integrand evaluations. In section~\ref{sec.examples} we illustrate the efficiency of our $hp$-quadrature algorithm on various examples of triangle, box and pentagon integrals in Euclidean kinematics, and we also provide a proof-of-concept that the method can be extended to physical scattering regions.  
In section~\ref{sec.conclusion} we present our conclusions. We include a set of appendices where we discuss some technical details about the algorithm omitted in the main text.

\section{One-loop integrals and Loop-Tree Duality}
\label{sec.LTD}

\subsection{Definitions}
We consider one-loop Feynman integrals in $D$ dimensions of the form
\begin{align}\label{feynman}
I=i\int  \frac{\rd^D k}{\pi^{\frac{D}{2}}} \frac{\mathcal{N}}{D_1\dots D_{N_P}}\,,
\end{align}
where in $D=4$ dimensions the integration domain is the four-dimensional real space with the usual (Lebesgue) measure. 
The $N_P$ propagators are denoted by
\begin{align}\label{feynprops}
D_i=q_i^2-m_i^2+i\epsilon\,,\quad q_i= k+\sum_{j=1}^{E-1}b_{ij}p_j\,,\quad b_{ij}\in\{0,\pm 1\}\,.
\end{align} 
Since the primary goal of our paper is to introduce a novel numerical method to evaluate one-loop integrals, we assume that the integral in eq.~\eqref{feynman} is convergent (we will comment on divergent cases below).
The integral in eq.~\eqref{feynman} is a function of the $N_P$ propagator masses $m_i^2\geq 0$ and the external momenta $p_j$, $1\le j\le E$ (note that at one loop, $E=N_P$), which are $D$-dimensional\footnote{For now, we treat the space-time dimension as an integer. We will comment on dimensional regularisation below.} vectors constrained by momentum conservation $\sum_{j=1}^Ep_j=0$. By Lorentz-invariance, the integral only depends on the external momenta through the Minkowski scalar products (of signature $(1,D-1)$) between external momenta,
\beq
p_i\cdot p_j = p_i^{\scriptstyle(0)}\,p_j^{\scriptstyle(0)} - \bs{p}_i\cdot \bs{p}_j\,,
\eeq
where $\bs{p}_i = (p_i^{\scriptstyle(1)},\ldots,p_i^{\scriptstyle(D-1)})$ are their spatial components. The numerator $\mathcal{N}$ in eq.~\eqref{feynman} is a polynomial in scalar products between the loop momentum $k$ and the external momenta. Finally, $\epsilon$ is an infinitesimal parameter that shifts the location of the poles of the propagators into the complex plane. 

One-loop integrals have been extensively studied in the literature. In particular, it is known that one-loop integrals can always be expressed in terms of polylogarithms~\cite{Mork:2026wie}. Systematic approaches to obtain analytic expressions in terms of polylogarithms are in principle available~\cite{Ren:2023tuj,Duhr:2025azh,Mork:2026wie}, though the resulting analytic expressions may be hard to evaluate. In $D=4$ dimensions only integrals with at most four propagators are relevant for the computation of one-loop amplitudes~\cite{Melrose:1965kb}, and using tensor reduction~\cite{Passarino:1978jh}, it is always possible to identify a basis of one-loop integrals with trivial numerator. 

\subsection{Divergencies}\label{subsec:div}

It is well known that one-loop integrals may diverge, and the divergences may be of ultraviolet (UV) or infrared (IR), i.e., soft or collinear, origin. Divergent integrals need to be suitably regularised before (numerical) evaluation. The most commonly used regularisation scheme is \emph{dimensional regularisation}~\cite{tHooft:1972tcz,Cicuta:1972jf,Bollini:1972ui}. In a nutshell, dimensionally-regulated loop integrals are evaluated in an arbitrary number $D=D_{\eps}$ of dimensions (for an axiomatic approach or an approach via twisted de Rham theory, see, e.g., refs.~\cite{Wilson:1972cf,Mastrolia:2018uzb}). We will work in a \mbox{`t Hooft}-Veltman (HV)-like scheme where the external momenta lie in four dimensions and only the loop momentum can access the $D_{\eps}-4$ remaining dimensions. The dimensionally-regulated integral then defines a meromorphic function of the complex variable $D_{\eps}$, which we can expand into a Laurent series around $\eps=0$, where $\eps=\tfrac{4-D_{\eps}}{2}$ is the deviation from $D=4$ space-time dimensions.

In most cases, the location of the singularities in loop-momentum space can be characterised by the \emph{Landau equations}. Recently, an algorithm was introduced to identify a set of numerators that render a Feynman integral finite~\cite{Gambuti:2023eqh}. This allows one to identify a so-called \emph{finite} basis, which consists of one-loop integrals that are convergent in four dimensions (possibly up to simple integrals that are known analytically).
In a nutshell, the algorithm of ref.~\cite{Gambuti:2023eqh} proceeds as follows:
The location of the soft and collinear singularities of the integral are determined via the Landau equations and power counting. We then determine the most general  numerator in terms of scalar products between the loop momentum and external momenta that render the integral both UV and IR finite. To this effect, we write down an ansatz for the numerator as a polynomial in the scalar products and construct a linear system for the coefficients in the ansatz by requiring the numerator to vanish with the adequate power counting at the location of the Landau singularities. Not all coefficients are typically fixed, but we get a family of possible numerators that render the integral IR finite. They generate an ideal in  $\mathbb{C}[k_\perp\cdot k_\perp,k\cdot v_1,k\cdot v_2,k\cdot v_3]$,
where $(v_1,v_2,v_3)$ is a van Neerven–Vermaseren basis~\cite{vanNeerven:1983vr}, which can be expressed in terms of Gram determinants of the external momenta, and $k_\perp$ is the projection of the loop momentum onto the space orthogonal to the (four-dimensional) space spanned the external momenta. The numerators that render the integral finite are in the subset of this ideal consisting of numerators that render the integral UV finite by power-counting. One may then choose a basis of integrals where most of the integrals are convergent in $D=4$ dimensions (except possibly simple ones). At one loop, we can always find a finite basis for integrals with at least three propagators. For tadpole and bubble integrals, this is not possible, but these integrals are known in analytic form to all orders in the dimensional regulator $\eps$, cf.,~e.g.,~refs.~\cite{tHooft:1972tcz}. 

We note that, even if the integrals are convergent in $D=4$ dimensions, they define multi-valued functions of the propagator masses and the external momenta. In particular, they may develop non-zero imaginary parts, and the regions of parameter space where the imaginary part is non-zero are directly related to thresholds where some of the propagators go on-shell. In the Euclidean region, which contains the region where all scalar products between external momenta are negative (or more generally for kinematics below threshold), the integral is real.

\subsection{Loop-Tree Duality}
We now focus on integrals that converge for $\eps=0$, and we only discuss the evaluation of the integral in exactly $D=4$ dimensions. We will come back to the evaluation of the higher orders in the Laurent expansion in $\eps$ in section~\ref{sec:DimReg}.
It is natural to attempt a direct numerical evaluation in loop-momentum space for convergent integrals. Already at one-loop, this leads to four-dimensional integrals that need to be evaluated. It is, however, possible to reduce the dimensionality of the integral to three, as we will review in this section.

We fix the external momenta $p_i$, and we regard the integrand as a function of the integration variables $k^{(i)}$. We then focus on the integration over the energy component $k^{(0)}$ of the loop momentum $k$. The propagators have simple poles in $k^{(0)}$. Due to the $i \epsilon$ terms in eq.~\eqref{feynprops}, the location of  the poles is infinitesimally shifted into the complex plane, and there are no poles on the real line. After closing the integration contour in the upper (or lower) complex half-plane, the residue theorem can be applied, and the integration over $k^{(0)}$ can be performed analytically. This reduces the dimensionality of a one-loop integral from four to three. This is the fundamental mechanism behind the  Loop-Tree Duality (LTD)~\cite{Catani:2008xa}. 

Rather than discussing LTD in general, we focus on an example that illustrates the main aspects. We stress that this example is very simple and well known in the literature. We nevertheless discuss it here, as it allows us to introduce the concepts needed in subsequent sections.
We consider the one-loop triangle integral in four space-time dimensions,
\beq\bsp\label{introtriang}
I_\triangle&=i\int \frac{\rd^4{k}}{ \pi^2}\frac{1}{[(k+p_1)^2-m_1^2+i\epsilon][(k+p_1+p_2)^2-m_2^2+i\epsilon](k^2-m_3^2+i\epsilon)}\\
&=i\int \frac{\rd k^{(0)} \,\rd \bs{k}}{\pi^2}\frac{1}{\left[\left(k^{(0)}+p_1^{\scriptscriptstyle(0)}\right)^2-E^2_1\right]\left[\left(k^{(0)}+p^{\scriptscriptstyle(0)}_1+p^{\scriptscriptstyle(0)}_2\right)^2-E_2^2\right]\left[\left(k^{(0)}\right)^2-E_3^2\right]}\,,
\esp\eeq
where in the second line we made the integration over the energy component $k^{(0)}$ explicit, and we denote the Lebesgue measure in three-dimensional Euclidean space by $\rd \bs{k}=\rd^3 \bs{k}$. The $E_i$ in eq.~\eqref{introtriang} are functions of the integration variable $\bs{k}$:
\begin{align}\label{eqens}
E_1&= \sqrt{\big(\bs{k}+\bs{p}_1\big)^2+m_1^2}\,,\quad E_2=\sqrt{\big(\bs{k}+\bs{p}_1+\bs{p}_2\big)^2+m_2^2}\,,\quad E_3= \sqrt{\bs{k}^2+m_3^2}\,.
\end{align}
We perform the integration over the energy component $k^{(0)}$ by closing the contour in the upper half-plane and taking residues around the poles of the propagators. We find
\begin{align}\label{triangleLTD}
I_\triangle&= \int \frac{2\rd\bs{k}}{\pi}\, \Bigg\{ \frac{1}{{E}_1 \big[\big(E_{1}^{\scriptscriptstyle(-,0)}\big)^2-{E}_3^2\big]\big[\big(E_{1}^{\scriptscriptstyle(0,+)}\big)^2-E_2^2\big]}\\
&+\frac{1}{E_2 \big[\big(E_{2}^{\scriptscriptstyle(0,-)}\big)^2-E_1^2\big] \big[\big( E_{2}^{\scriptscriptstyle(-,-)}\big)^2-E_3^2\big]}+\frac{1}{{E}_3 \big[\big(E_{3}^{\scriptscriptstyle(+,0)}\big)^2-E_1^2\big] \big[\big(E_{3}^{\scriptscriptstyle(+,+)}\big)^2-{E}_2^2\big]}
\Bigg\}\notag\,,
\end{align}
where we defined
\begin{align}\label{trianglesurfaces}
E_{3}^{\scriptscriptstyle(+,+)} &:= E_3 + p_1^{\scriptscriptstyle(0)} + p_2^{\scriptscriptstyle(0)}\,, 
&\quad E_{2}^{\scriptscriptstyle(-,-)} &:= E_2 - p_1^{\scriptscriptstyle(0)} - p_2^{\scriptscriptstyle(0)}\,, 
&\quad E_{2}^{\scriptscriptstyle(0,-)} &:= E_2 - p_2^{\scriptscriptstyle(0)}\,, \\
E_{1}^{\scriptscriptstyle(0,+)} &:= E_1 + p_2^{\scriptscriptstyle(0)}\,, 
&\quad E_{1}^{\scriptscriptstyle(-,0)} &:= E_1 - p_1^{\scriptscriptstyle(0)}\,, 
&\quad E_{3}^{\scriptscriptstyle(+,0)} &:= E_3 + p_1^{\scriptscriptstyle(0)}\,.\notag
\end{align}
Equation~\eqref{triangleLTD} is called the \emph{LTD representation} of the triangle integral $I_\triangle$, and
we call the integrand in eq.~\eqref{triangleLTD} the \textit{LTD integrand}. We briefly discuss some of the properties of the LTD integrand, in particular its singularity structure.

The singularities of the LTD representation are governed by the denominators in the LTD integrand, which can be cast in a factorised form as 
\begin{align}\label{eq:EE_factored_HE}
\big(E_{i}^{\scriptscriptstyle(p,q)}\big)^2-E_j^2=\big(E_{i}^{\scriptscriptstyle(p,q)}-E_j\big)\big(E_{i}^{\scriptscriptstyle(p,q)}+E_j\big)\,.
\end{align}
We may then study the singular loci of the LTD integrand, i.e., the loci in three-dimensional loop-momentum space where the LTD integrand is singular. We obtain two distinct classes of singular surfaces, corresponding to which of the two factors in the right-hand side of eq.~\eqref{eq:EE_factored_HE} vanishes.
The equation $E_{i}^{\scriptscriptstyle(p,q)}-E_j=0$ describes a hyperbolic surface in three-dimensional space, referred to as an \emph{H-surface}. Instead, the equation  $E_{i}^{\scriptscriptstyle(p,q)}+E_j=0$ defines an ellipsoid called an \emph{E-surface}. We thus see that the propagator singularities of an LTD integrand of a (convergent) integral can be associated to either $H$- or $E$-surfaces. We note here a topological distinction between $E$- and $H$-surfaces: $E$-surface are ellipsoids, and thus compact, while the $H$-surfaces are non-compact.

It can be shown that the singularities associated to $H$-surfaces are spurious, and they cancel locally and pairwise in loop-momentum space. This mechanism is referred to as \textit{dual cancellation} in refs.~\cite{Aguilera-Verdugo:2019kbz,Buchta:2015wna,Buchta:2015xda}. These spurious singularities may render the numerical integration intricate, and the dual cancellations may cause numerical instabilities. It is therefore convenient to find a variant of the LTD representation where the dual cancellations are manifest and all singularities associated to $H$-surfaces are absent. It was shown in ref.~\cite{Capatti:2019ypt} that it is always possible to find a representation without spurious singularities that agrees with the original LTD integrand point-by-point in three-dimensional loop-momentum space. This representation is called \textit{manifestly causal Loop-Tree Duality} (cLTD) (see also refs.~\cite{Aguilera-Verdugo:2020set,Kromin:2022txz}), and its integrand is then purely expressible in terms of the energies and functions that are singular on $E$-surfaces. Each $E$-surface  corresponds to two propagators $D_i$ and $D_j$ going on-shell, and $D_i$ is the propagator around which the residue was taken in the derivation of the LTD integrand. At one loop, the corresponding terms admit the general form
\begin{align}\label{Esurfs}
E_{ij} &:=  E_i+E_j - c_{i,j} \,,
\end{align}
with 
\begin{align}\label{defC}
c_{i,j}:=%
\sum_{k=1}^{E-1}b_{jk}p^{(0)}_k-\sum_{l=1}^{E-1}b_{il}p^{(0)}_l\,.
\end{align}

For the triangle integral in eq.~\eqref{triangleLTD}, the cLTD representation can easily be obtained by partial fractionating,
\begin{align}\label{LTDTriangleCausal}
I_\triangle&=\int \frac{2\rd\bs{k}}{\pi}\frac{1}{4 E_1 E_2 E_3}\Bigg[\frac{1}{(E_{1}^{\scriptscriptstyle(-,0)}+E_3) (E_{2}^{\scriptscriptstyle(-,-)}+E_3)}\notag+\frac{1}{(E_{2}^{\scriptscriptstyle(0,-)}+E_1) (E_{2}^{\scriptscriptstyle(-,-)}+E_3)}\\
&+\frac{1}{(E_{1}^{\scriptscriptstyle(0,+)}+E_2) (E_{1}^{\scriptscriptstyle(-,0)}+E_3)}+\frac{1}{(E_{3}^{\scriptscriptstyle(+,0)}+E_1) (E_{3}^{\scriptscriptstyle(+,+)}+E_2)}+\frac{1}{(E_{1}^{\scriptscriptstyle(0,+)}+E_2) (E_{3}^{\scriptscriptstyle(+,+)}+E_2)}\notag\\
&+\frac{1}{(E_{2}^{\scriptscriptstyle(0,-)}+E_1) (E_{3}^{\scriptscriptstyle(+,0)}+E_1)}\Bigg]\notag\\
&=\int \frac{2\rd\bs{k}}{\pi}\frac{1}{4 E_1 E_2 E_3}\Bigg[\frac{1}{E_{31} E_{32}}+\frac{1}{E_{21}E_{31}}+\frac{1}{E_{13} E_{23}}+\frac{1}{E_{21}E_{23}}+\frac{1}{E_{12} E_{13}}\Bigg].
\end{align}
As expected, the only denominators in the integrand in eq.~\eqref{LTDTriangleCausal} are those that define the $E$-surfaces. For arbitrary LTD integrands -- possibly including additional numerators -- the systematic cancellation of spurious $H$-surfaces is detailed in ref.~\cite{Capatti:2020ytd}. 
Note that there exist other representations that are free of singularities lying on $H$-surfaces, such as the \textit{cross-free family} representation~\cite{Capatti:2022mly}, which can be constructed  diagrammatically. 
In the one-loop case, ref.~\cite{Kromin:2022txz} formalizes an alternative explicit formula for the cLTD integrand along with a recursion relation, and it additionally offers a graphical representation of the cLTD integrand. We note that a main challenge when working with the cLTD representation can be the exponentially growing number of terms with the number of propagators, which may result in very large expressions for the cLTD integrand, especially at higher loop orders or for integrals depending on many external momenta.

 \subsection{One-loop integrals to higher orders in the dimensional regulator}
 \label{sec:DimReg}

 In dimensional regularisation, explicit analytic results are available for all one-loop integrals with at most four propagators in $D_{\eps}=4-2\eps$ dimensions, up to terms that vanish as $\eps\to0$, cf.,~e.g.,~refs.~\cite{tHooft:1972tcz,Ellis:2007qk,Carrazza:2016gav,Barik:2026twt}. This is sufficient for all phenomenologically-relevant one-loop computations. Higher-order computations, however, do not only involve the evaluation of loop integrals with more than one loop, but they also receive contributions from one-loop integrals evaluated to higher orders in the dimensional regulator, including integrals depending on more than four propagators.
In a HV-like scheme, we may restrict to integrals depending on at most five propagators (cf., e.g., ref.~\cite{Bern:1993kr}). While analytic results in terms of polylogarithms also exist for those cases, they are much more complicated, and a direct numerical approach may be desirable. A direct application of the LTD method, however, needs to deal with the fact that we need to perform numerical integration in an integer number of dimensions. We now review a method that allows one to cast the higher orders in the expansion around $\eps=0$ in the form of onefold integrals over the corresponding integrals in strictly four dimensions. 

The starting point is the realisation that in a HV-like scheme, we may decompose the $D_{\eps}$-dimensional loop momentum $k$ as 
\beq
k = k_{\parallel} + k_{\perp}\,,
\eeq
where $k_{\parallel}$ denotes the orthogonal projection of the loop momentum onto the four-dimensional space in which the external momenta lie, and $k_{\perp}$ is the projection onto the orthogonal complement. The van Neerven-Vermaseren basis introduced in subsection~\ref{subsec:div} is one way to achieve such a decomposition. By construction, it satisfies
\beq
k\cdot p_i = k_{\parallel}\cdot p_i \textrm{~~~and~~~} k_{\parallel}\cdot k_{\perp} = k_{\perp}\cdot p_i =0\,, 
\quad 1\le i\le E\,.
\eeq
The $D_{\eps}$-dimensional integral can then be cast in the form:
\begin{align}
\int \rd^{D_{\eps}} f(k\cdot p_i, k \cdot k,m_j^2)&=\int \rd^4 k_{||}\, \rd^{-2 \eps}k_\perp f(k\cdot p_i, k^2 ,m_j^2)\notag \\
&=S_{-2\eps-1}\int_0^\infty \rd k_\perp\,k_\perp^{-2\eps-1}\int \rd^4 k_{||}f(k\cdot p_i, k^2,m_j^2)\\
&=\frac{\pi^{-\eps}}{\Gamma\left(-\eps\right)}\int_0^\infty \rd\eta \,\eta^{-\eps-1}\int d^4 k_{||}f(k\cdot p_i, k\cdot k,m_j^2)\,,\notag
\end{align}
 and $S_{-2\eps-1}$ is the area of the $(-2\eps-1)$- dimensional unit sphere\footnote{Note that for $n\in \mathbb{N}_+$ we have $\int \rd^n x=\int|x|^{n-1}\rd |x| \rd \Omega_n=S_{n-1}\int|x|^{n-1}\rd |x|=\frac{2 \pi^{n/2}}{\Gamma(n/2)}\int  |x|^{n-1}\rd|x|$, where $S_{n-1}$ is the area of the $n-1$-dimensional unit sphere. Then for $n=-2\eps$ this is defined by analytic continuation and the solid angle becomes $S_{-2\eps-1}=\frac{2 \pi^{-\eps}}{\Gamma(-\eps)}$.} and $\eta=|k_\perp|^2$. The function $f(k\cdot p_i, k^2, m_j^2)$ is a product of propagators, and we find 
\begin{equation}\begin{split}
f(k\cdot p_j,  k^2,m_i^2)&\,=\prod_i\frac{1}{\left(k+\sum_{j=1}^{E-1}b_{ij}p_j\right)^2-m_i^2}\\
&\,=\prod_i\frac{1}{\left(k_{\parallel}+\sum_{j=1}^{E-1}b_{ij}p_j\right)^2-(m_i^2+\eta)}\\
&\,=f(k_{\parallel}\cdot p_j,  k_{\parallel}^2,m_i^2+\eta)\,,
\end{split}\end{equation}
with $b_{ij}$ as in eq~\eqref{feynprops}.
We see that the $D_{\eps}$-dimensional integral can be written as a onefold integral over the four-dimensional integral,
\begin{align}\label{higherordersgen}
I^{D_{\eps}}=\frac{1 }{\Gamma(-\eps)}\int_0^\infty \rd \eta \,\eta^{-1-\eps} I_{|m_i^2\rightarrow m_i^2+\eta}\,.
\end{align}
The integral over $\eta$ diverges in the limit $\eps\to0$ and produces a pole in $\eps$ which cancels against the $\Gamma$ function in the prefactor. We then first need to subtract the singularity at $\eta=0$ before we can perform the integration. This is, however, merely a technical complication, and we refer to,~e.g.,~ref.~\cite{Duhr:2025azh} for more details. 
The remaining factor in the integrand is the integral in $D=4$ dimensions, with all propagator masses shifted by $\eta$. After subtracting the singularity at $\eta=0$, we may perform the integration order by order in $\eps$. In particular, we may replace four-dimensional integrals by their (c)LTD representations, and we obtain in this way a four-dimensional (c)LTD-like representation for the higher orders in the $\eps$-expansion of one-loop integrals.

\subsection{Summary}By combining the results from the previous section, we obtain an effective algorithm to evaluate one-loop integrals in $D_{\eps} = 4-2\eps$ numerically:

\begin{enumerate}
    \item We follow the prescription of ref.~\cite{Gambuti:2023eqh} and we choose a basis of one-loop integrals with numerators that render the integrals convergent in four dimensions. At one loop, this can always be achieved for integrals with at least three propagators. 
\item  We write the integrals in $D_{\eps}$ dimensions as onefold integrals over their four-dimensional counterparts as in eq.~\eqref{higherordersgen}. 
\item We derive a cLTD representation for the four-dimensional integrals. As a consequence, for every order in $\eps$, we obtain a (at most) four-dimensional integral representation. By construction, this integral is convergent in Euclidean kinematics, where there are no threshold singularities. It can thus be directly evaluated  using numerical algorithms.
\end{enumerate}

The standard tool in (high-energy) physics to evaluate multi-dimensional integrals numerically is MC integration. The convergence of the MC integration scales with the inverse of the square root of the number of points in the root mean square error. In particular, the convergence is independent of the dimensionality of the integral. While it is in principle possible to achieve any desired accuracy with MC integration by evaluating the integrals at enough points, the precision that can be achieved in practice is typically limited by the rather slow convergence of the algorithm, leading to a trade-off between precision and runtime in applications. Moreover, since MC integration relies on random sampling, its results are inherently stochastic and can therefore, with decreasing probability with the number of samples, deviate from the true solution by an arbitrary amount. In the remainder of this paper, we argue that for the specific integral representations derived in this section, there is an alternative, deterministic algorithm, based on quadrature rules with (stretched) exponential convergence.

\section{$hp$-Quadrature}
\label{sec.hp}
In this section we present the main result of our paper, and we introduce a novel algorithm based on quadrature rules particularly well suited to evaluate the cLTD-type representations of one-loop integrals from the previous section. We focus in this section on the mathematical background, and we will discuss various exemplary evaluations of one-loop integrals in the next section. We start by giving a general review of the method (mainly intended for physicists), but we also present the concrete mathematical theorems that allow us to have a rigorous mathematical bound on the error estimate and the convergence of this algorithm.

\subsection{Summary of the algorithm}

We start by giving a pedestrian overview of quadrature rules for numerical integration in general and our algorithm in particular, focusing on why this algorithms is well adapted to the evaluation of one-loop integrals in the cLTD representation.

\subsubsection{Numerical integration via quadrature rules}

The idea of numerical quadrature is to approximate an integral by a weighted mean of point evaluations of a continuous integrand,
\beq
\int_H f(x)\,\mathrm{d}x
\approx
\sum_{i=1}^N\omega_if(\xi_i)\,,
\eeq
where $\{\xi_i\}_{i=1}^N$ and $\{\omega_i\}_{i=1}^N$ are the quadrature nodes and weights, and $H$ is the integration domain. For simplicity we restrict ourselves to the well-known Gauss-Legendre quadrature rules. Other quadrature rules can be used with minor modifications.

Let us start by reviewing Gauss-Legendre quadrature rules for one-dimensional integrals. On the unit interval $H:=[-1,1]$, the quadrature nodes of the $N$-point quadrature rule are given as the roots of the Legendre polynomial of degree $N$ and the weights are the integrals of the corresponding Lagrange polynomials. We refer to ref.~\cite{AS1964} for a table of Gauss-Legendre nodes on $[-1,1]$. The generalisation to other intervals $[a,b]$ is done by a suitable affine scaling. We denote the $N$-point Gauss-Legendre quadrature rule applied to a continuous function $f$ on $[a,b]$ by $\mathcal{Q}_N^{[a,b]}[f]$.

Let us now turn to the extension of Gauss-Legendre quadrature to higher dimensions.
We consider an axis-parallel $d$-dimensional hypercube in Euclidean space,
\begin{equation}\label{eq:cube}
H := \bigtimes_{i \in [d]}[a_i, a_i+h]\,,
\end{equation}
where $[d] := \{1,\dots,d\}$, $a_i \in \mathbb{R}$ and $h \in \mathbb{R}_{>0}$. Taking the tensor product of one-dimensional Gauss-Legendre quadratures yields a $d$-dimensional quadrature rule
\beq
\mathcal{Q}^{H}_{N} [f]
:=
\left( \bigotimes_{i\in[d]} \mathcal{Q}^{[a_i, a_i+h]}_{N} \right)[f]
:=
\sum_{i_1\in[N]}\cdots\sum_{i_d\in[N]} 
f(\xi_{i_1},\ldots,\xi_{i_d})w_{i_1}\cdots w_{i_d}\,,
\eeq
where $\{\xi_i\}_{i=1}^N$ and $\{\omega_i\}_{i=1}^N$ are the nodes and weights of the Gauss-Legendre quadrature. The idea is that $\mathcal{Q}^{H}_{N} [f]$ approximates the $d$-dimensional integral $\int_H f({\bf x})\,\rd {\bf x}$. The $N$-point Gauss-Legendre rules are exact for every polynomial of degree at most $2N-1$, i.e.,
\beq
\mathcal Q_N^{H}[p]
=
\int_H p(\mathbf{x})\,\rd\mathbf{x}\,,
\eeq
for all polynomials $p$ with $\deg(p)\le 2N-1$. It is known that this is optimal in terms of polynomial exactness.

Suppose now that $f$ is not a polynomial, but merely continuous. The idea is to pick a polynomial that approximates $f$ and that can be integrated exactly using the quadrature rule. In particular, if $f$ is continuous, then there exists a unique
interpolation polynomial $p$ of degree at most $N-1$ passing through the values of $f$ at the quadrature nodes. Since the quadrature rule integrates this polynomial exactly,
\beq
\mathcal Q_N^H[f]
=
\mathcal Q_N^H[p]
=
\int_H p(\mathbf{x})\,\rd\mathbf{x}\,,
\eeq
the quadrature error is entirely determined by how well $p$ approximates $f$.
We note that the existence of polynomial approximations is guaranteed by the Stone--Weierstrass theorem, which states that every continuous function on a compact domain can be approximated arbitrarily well by polynomials. However, this theorem does not provide any information on the required polynomial degree.

A much stronger statement holds whenever $f$ is analytic. Assume that $f$ extends holomorphically to a complex polydisk  of radius $r$ around a point $\mathbf{z}_0$,
\begin{equation}
    P_{\mathbf{z}_0}(r) = \{\mathbf{z}=(z_1,\dots,z_d) \in \mathbb{C}^d : \;|z_i-z_{0,i}| < r, \quad \text{for all }i \in [d]\}\,.
\end{equation}
If $f$ is holomorphic on $D\subset\mathbb{C}^d$, it is also analytic on $D$, i.e., for every point $\mathbf{z}_0\in D$ there is a polydisk $P_{\mathbf{z}_0}(r)\subset D$ with $r>0$ for which the Taylor series 
\beq 
\sum_{\bs{\nu}} c_{\bs{\nu}} (\mathbf{z}-\mathbf{z}_0)^{\bs{\nu}}\,,\qquad\bs{\nu} = (\nu_1,\ldots,\nu_d)\,,\qquad  \mathbf{z}^{\bs{\nu}} = \prod_{i=1}^dz_i^{\nu_i}\,,
\eeq
on  $P_{\mathbf{z}_0}(r)$ converges to $f$. Here the Taylor coefficients are 
\beq
c_{\bs{\nu}}=\frac{\partial^{\bs{\nu}f}(\mathbf{z}_0)}{\boldsymbol{\nu}!}\,, \qquad \boldsymbol{\nu}! := \nu_1! \dots \nu_d!\,.
\eeq
Cauchy's estimate implies that there exists a $C_f>0$ such that
\beq\label{eq:Cauchy_estimate}
|c_{\bs{\nu}}|
\le
\frac{C_f}{r^{\|\boldsymbol{\nu}\|_1}}\,,
\qquad
\|\boldsymbol{\nu}\|_1
=
\nu_1+\ldots+\nu_d\,.
\eeq
Hence, the Taylor coefficients decay rapidly, showing that analytic functions are already well approximated by relatively low-degree polynomials. Consequently, a quadrature rule that is exact for polynomials of sufficiently low degree is expected to provide an accurate approximation of the integral. In particular, increasing the distance $r$ to the nearest singularity further improves the polynomial approximation and therefore the convergence of the quadrature rule.

By Cauchy's estimate in eq.~\eqref{eq:Cauchy_estimate}, there exists a $C_f\in \mathbb{R}_{\geq 0}$ for a hypercube $H$ such that we have the following bound:
 \begin{align}\label{eq:L_infty_bound}
    \|\,\partial^{\boldsymbol{\nu}} f \|_{C({H})} \leq \frac{C_f}{r^{\|\boldsymbol{\nu}\|_1}} \boldsymbol{\nu}!\,, \quad \forall \boldsymbol{\nu} \in \mathbb{N}^d_0\,.
    \end{align}
Here, $\|\cdot \|_{C({H})}$ denotes the supremum norm for continuous functions, which is defined as
\begin{align}
\|f\|_{C(H)} := \operatorname*{sup}_{\mathbf{x}\in H}|f(\mathbf{x})|\,.
\end{align}
As explained above, \(r\) is the radius of a polydisk centered around the point $\mathbf{z}_0\in H$ that is closest to a singularity. Determining these polydisks directly may be difficult. It is therefore convenient to derive a slightly weaker estimate than eq.~\eqref{eq:L_infty_bound} using the Euclidean distance to the singularity set $S$. Let us define 
     \begin{equation}\label{eq:dist_def}
        \operatorname{dist}(A,B) := \inf_{\mathbf{a} \in A} \inf_{\mathbf{b} \in B}  |\mathbf{a}-\mathbf{b}|\,.
        \end{equation}
By definition, there are no singularities in the Euclidean ball with radius $r<\operatorname{dist}(H,S)$ and centered at any $\bs{z}_0\in H$. Then we realize that into any $d$-sphere of radius $r$ we can fit  a polydisk of radius $\frac{r}{\sqrt{d}}$. Hence the following estimate holds
    \begin{align} \label{eqn:analytic_cons_propsoverview}
    \|\ \partial^{\boldsymbol{\nu}} f \|_{C({H})} \leq \frac{C_f}{\gamma^{\|\boldsymbol{\nu}\|_1}} \boldsymbol{\nu}!\,, \quad \forall \boldsymbol{\nu} \in \mathbb{N}^d_0\,,\quad \text{ with }\quad \gamma=\frac{\mathrm{dist}(H,S)}{\sqrt{d}}\,.
    \end{align}
One can show (as a direct consequence of ref.~\cite{boerm2010efficient}, Theorem 4.16, and Hölder's inequality) that the error of the $N$-point Gauss-Legendre quadrature on $H$, with $H$ as in \cref{eq:cube}, is bounded as
    \beq\label{eq:GLerroroverview}
    \left| \int_H f(\mathbf{x})\, \rd\mathbf{x} - \mathcal{Q}^H_N[f]\right| \leq C h^d \left(1+\frac{h}{\gamma}\right)\left( \frac{h}{2\gamma + \sqrt{4h^2+\gamma^2}} \right)^{N}\,,
    \eeq
where $C$ depends on $C_f$, $N$ and $d$ as in eq.~\eqref{eq:cube}.
We can now clearly see that the Gauss-Legendre quadrature error for a function analytic on ${H}$ scales algebraically with the side length $h$ and the derivative growth constant $\gamma$ from \cref{eqn:analytic_cons_propsoverview}. The error decay is exponential with respect to the number of point evaluations $N$.

\subsubsection{$hp$-quadrature}\label{sec:history}

As introduced above, our analysis relies on rigorous error estimates for Gauss–Legendre quadrature of analytic integrands. Although the cLTD integrand is not globally analytic, it is analytic away from a compact set of singularities. This local analytic structure makes cLTD particularly well suited to an \(hp\)-quadrature strategy: $hp$-quadrature is a specific type of quadrature which constructs weights and nodes by means of a combination of simpler quadrature rules of varying order on a partition of the integration domain. The partition and quadrature order are determined using a priori knowledge of the singularities of the integrand.  More explicitly, we construct a mesh of hypercubes that avoids the singular set but may approach it arbitrarily closely. The size of each hypercube and the corresponding quadrature order are then chosen according to its distance from the singular set. 

The method achieves stretched-exponential convergence rates, i.e., a decrease of the approximation error like $\mathcal{O}(\exp(-bE^a))$ in the number of function evaluations $E$ for some $a,b>0$, when these singularities are point singularities, as we will prove below. This is the case for one-loop cLTD integrands the Euclidean region. We expect the method more generally to be well-suited also for the (compact) singularity structure of also integrals beyond the Euclidean region.

$hp$-quadrature is the sophisticated combination of quadrature rules according to an $hp$-refinement strategy and has emerged as a combination of $h$-refinement and $p$-refinement, both of which can be motivated from \cref{eq:GLerroroverview}.
$h$-refinement fixes a quadrature rule and subdivides the domain of integration into subdomains of equal size to employ the same quadrature rule on each of the subdomains. The algebraic scaling of the quadrature error in \cref{eq:GLerroroverview} yields convergence of this composite quadrature rule in $h$, with only $N+1$ well-behaved derivatives required when an $N$-point quadrature rule is used. This moderate smoothness requirement comes at the price of an algebraic convergence at rate.
In contrast, $p$-refinement does not subdivide the integration domain, but aims at increasing the order of the quadrature rule by increasing the number of points $N$ in the quadrature rule. For polynomial-based quadrature rules such as Gauss-Legendre rules, this is achieved by increasing the polynomial degree, hence $p$-refinement. As is illustrated by \cref{eq:GLerroroverview}, this yields stretched-exponential convergence, provided the integrand is analytic and with the contraction rate depending on the constants in \cref{eqn:analytic_cons_propsoverview}. We refer to the relevant textbooks such as refs.~\cite{Gau2012,SB1993} for more details on $h$- and $p$-refinement. Summarized, for analytic integrands, $p$-refinement can have a faster convergence and better cost-to-accuracy ratio. For integrands of lower regularity, $h$-refinement is often the only reliable alternative.

The motivation for $hp$-refinement is to combine the best of the two worlds of $h$- and $p$-refinement to obtain efficient quadrature methods even for locally non-smooth integrands. The explicit combination of the two refinement mechanisms emerged originally from the $hp$-finite-element literature. Early structural-mechanics work in the late 1970s had already explored locally varying polynomial orders and self-adaptive high-order elements~\cite{PEANO197869}, while from the early 1980s~\cite{babuvska1981error} to the middle 1990s the $hp$-version had become a systematic framework in which local mesh refinement and local polynomial enrichment were treated jointly~\cite{BABUSKA19905}.
From a methodological perspective, the main ingredients of the, to be presented, $hp$-refinement strategy are well established and have been studied extensively~\cite{refId0, Schwab1994, Boerm2005}, although in a geometrically simpler setting than required for cLTD integrands. The takeaway from these works is to notice that a-priori knowledge about regularity and singularity sets of integrands permit explicit refinement and convergence estimates. We show in the following that such a-priori knowledge can be obtained for cLTD integrands and that $hp$-refinement is indeed applicable with theoretically guaranteed stretched-exponential convergence rates.

\subsubsection{$hp$-quadrature and cLTD for one-loop integrals}

In this section we motivate why $hp$-quadrature is well suited to evaluate cLTD representations of one-loop integrals. We only give a rough overview of the idea. Detailed mathematical proofs will be presented in section \ref{sec:hp-quadrature}. 
In particular, we explain why the Gauss-Legendre quadrature cannot naively be applied to the cLTD integrand. We describe the relevant complications and then show how $hp$-quadrature provides a systematic strategy for addressing them.

For simplicity, and as a proof of concept, we focus here on Euclidean kinematics (or more generally, kinematics below threshold, where the integral is real), and we leave the discussion of how to extend the method to include the imaginary part generated when crossing thresholds to future work~\cite{hp-quad-imag}. We also focus on finite integrals in $d=4$ dimensions. Extending the method to include higher orders in $\eps$ does not constitute any bottleneck, and we will come back to this in section~\ref{sec.examples}.

\paragraph{Singular structure of the cLTD integrand.} Assume that we want to evaluate a one-loop Feynman integral using a Gauss-Legendre quadrature rule. The integrands of Feynman integrals are never holomorphic or analytic everywhere, so that the error analysis of plain Gauss-Legendre quadrature does not apply, and the quadrature error is not predictable.

If we restrict to the Euclidean kinematic region and assume that the corresponding integral is finite, we may take the limit $\epsilon\to0$ in the Feynman $i\epsilon$ prescription. 
Even for finite integrals in the Euclidean region, the cLTD integrand is not analytic on $\mathbb{C}^3$ (nor is it real-analytic on $\mathbb{R}^3$). In particular, the overall prefactor of the cLTD integrand contains inverse powers of the internal energies,
\begin{align}\label{eq:sqrt_def}
E_i=\sqrt{\bs{q}_i^{2}+m_i^2}\,,
\end{align}
with $\bs{q}_i$ defined in eq.~\eqref{feynprops}, see for example eq.~\eqref{LTDTriangleCausal}. The intersection of the singular loci of the $E_i$ with the real integration domain consists of isolated points (see the discussion below).
Although these point singularities are integrable, they obstruct the analyticity of the integrand, and we need to include them into the set $S$ of singularities, understood as the set in $\mathbb{C}^n$ where the integrand is not analytic.
We note here that differentiating terms containing singularities associated with $E$-surfaces introduces additional inverse powers of $E_i$. Higher derivatives therefore enhance the singular behaviour at the same branch points of the energies.\footnote{It is known that the integrand can generally be rewritten such that the  $E_i=0$ singularities are removed. This is referred to as \textit{multi-channeling} (cf., e.g., ref.~\cite{Capatti:2019edf}).  However, even after multi-channeling higher derivatives of the individual terms  reintroduce singularities at the exact same branch points. Therefore we  keep these singularities as they are.}  

The locus of singularities $S$ can be written as a union $S=S_F\cup S_E$:
\begin{itemize}
\item $S_E$ contains the singularities originating from the $E$-surfaces. In the Euclidean region, the
the $E$-surfaces never intersect the real integration domain, $S_E\cap\mathbb{R}^3=\emptyset$.  
\item $S_F$ is the set of branch points of the energies, cf.~eq.~\eqref{eq:sqrt_def}. In contrast to the $E$-surfaces, its intersection with the real domain is not empty even in the Euclidean region, and we define $P:=S_F\cap\mathbb{R}^3$ as the set of branch points on the integration contour. 
\end{itemize}
In the cLTD representation these are the only possible singularities of the integrand. Note that both classes of singularities are confined to compact regions, so that the union $S$ is also compact. 
For later convenience, we write\footnote{Note that here $\mathbf{z}\cdot\mathbf{z} = \sum_{i=1}^dz_i^2$ is the real inner product applied to complex vectors.}
\begin{equation}\label{eq:SESF}
S_F= \bigcup_{i \in[N_P]} S_{F,i}\,,\quad \text{ with }\quad S_{F,i} = \left\{ \mathbf{z} \in \mathbb{C}^d : (\mathbf{z}-\boldsymbol{\alpha}_i)\cdot (\mathbf{z}-\boldsymbol{\alpha}_i)+m_i^2\right\}\,,
\end{equation} 
where, as before, $N_P$ is the number of propagators. Further, $m_i\in \mathbb{R}_{\geq0}$ is the mass of the $i^{\textrm{th}}$ propagator and $\bs{\alpha}_i:=\sum_{j=1}^{E-1}b_{ij}\bs{p}_j$ is the spatial part of the momentum flowing through the $i^{\textrm{th}}$ propagator, with $b_{ij}$ as in eq.~\eqref{feynprops}. We write $\mathbf{z}$ instead of the loop momentum $\bs{k}$ to stress that $\mathbf{z}$ is a complex vector.
The intersection of $S_F$ with the real integration domain is the finite set of points,
\begin{align} \label{eqn:sing_point_set}
P = \{\boldsymbol{\alpha}_1,\ldots,\boldsymbol{\alpha}_{N_P} \}\,,\qquad \boldsymbol{\alpha}_i \in \mathbb{R}^d\,.
\end{align}
The $\boldsymbol{\alpha}_i$ are exactly the focal points of the ellipsoids defined by the $E$-surfaces, and so we will also refer to them as the \emph{focal points of the cLTD representation}.
We write 
\begin{align}\label{eq:SESF2}
S_E&=\bigcup_{(i,j) \in I}  S_{E,(i,j)}\,,\quad \text{ with } \quad I \subset [N_P] \times [N_P]\,,
\end{align}
where by $S_{E,(i,j)}$ we denote the complex zero set of the $E_{ij}$ defined in eq.~\eqref{Esurf}. Explicitly, these are given by
\begin{align}
S_{E,(i,j)} &= \left\{ \mathbf{z} \in \mathbb{C}^d : \sqrt{\|\mathbf{z} - \boldsymbol{\alpha}_{i} \|^2+m_i^2} + \sqrt{\|\mathbf{z} - \boldsymbol{\alpha}_{j} \|^2+m_j^2} - c_{i,j} = 0 \right\}\,,\notag
\end{align}
with $c_{i,j}$ as in eq.~\eqref{defC}.

\paragraph{Analytic structure of the cLTD integrand.}
In the following $f$ denotes a one-loop cLTD integrand, for example, $f$ may be the integrand in eq.~\eqref{LTDTriangleCausal}. 
Observe from \cref{eqens} that each energy \(E_i\) is given through
\beq
E_i(\mathbf{x})
:=
\sqrt{\|\mathbf{x}-\boldsymbol{\alpha}_i\|^2+m_i^2}\,,
\eeq
i.e., the square root of a quadratic polynomial. The argument of the square root is strictly positive for all $\mathbf{x}\in\mathbb{R}^3$, except when $m_i=0$ and $\mathbf{x}=\boldsymbol{\alpha}_i\in P$, with the set of focal points $P$ as in \cref{eqn:sing_point_set}. This makes $E_i$ an analytic function on $\mathbb{R}^3 \backslash P$. With the sum of anylytic functions being analytic, this makes the quantity $E_{ij}$
 in \cref{Esurfs} analytic on $\mathbb{R}^3$. By the assumption that the corresponding $E$-surfaces do not intersect the real integration domain, the \(E_{ij}\) are strictly positive, i.e., $E_{ij}>0$ on $\mathbb{R}^3$. This makes $1/E_{ij}$ analytic on $\mathbb{R}^3$ as well. 

With the product and sum of analytic functions on $\mathbb{R}^3 \backslash P$ being analytic, it is easy to see that $f$ satisfies the bound on derivatives given in \cref{eqn:analytic_cons_propsoverview} for some unknown positive $\gamma$.
We show in \cref{thm:general_holomorphic_derivative_estimate} that \cref{eqn:analytic_cons_propsoverview} is indeed satisfied, justifying the use of $hp$-quadrature for cLTD integrands.

\paragraph{Integration domain.} We now want to cover most of the integration domain $\mathbb{R}^3$ by differently sized hypercubes $H\subset \mathbb{R}^3$ on which $f$ is analytic. This implies in particular that $\gamma>0$. We then apply the Gauss-Legendre quadrature on these hypercubes.
As the hypercubes are compact, we will not be able to cover the full domain $\mathbb{R}^3$ by such cubes. A good choice is to truncate the total domain $\mathbb{R}^3$ to a hypercube 
\begin{align}\label{eq:defb}
D_T=[-b,b]^3\,,
\end{align}
with $b$ large enough such that it contains all singularities, i.e., $S\subset D_T$. Note that, since $S$ is compact, we can always find $D_T$. We comment on the exact choice of $b$ in the appendix~\ref{app:choiceofb}. The residual domain
\begin{align}\label{eq:defDR}
D_R=\mathbb{R}^3\backslash D_T
\end{align}
is free of singularities and
is integrated numerically with machine precision as described in appendix~\ref{app:resdom}. We then partition $D_T$ further into smaller hypercubes of different sizes. In particular these cubes never intersect or touch the singularities in $S_F$. As these are point singularities, $D_T\backslash S_F$ is not compact, and there will be again a small domain $R_U$ that is neglected and contributes to the total error.

\begin{thmblock}
\begin{lemma}\label{thm:quad_error_scaling}
Let \(f\) be an integrand of the form given in \cref{LTDTriangleCausal}, and assume that the E-surfaces do not intersect the real integration domain $S_E\cap\mathbb{R}^3=\emptyset .$

Let $G:=\mathbb{R}^3\setminus P,$
where $P$ are the focal points is given in \cref{eqn:sing_point_set}. Then, for every compact bounded open set \(G_D\subset G\), there exist constants \(C_f,\gamma>0\) such that
\begin{align}\label{eqn:analytic_cond_first}
\left\|\partial^{\boldsymbol{\nu}}f\right\|_{C(G_D)}
\leq
\frac{C_f}{\gamma^{\|\boldsymbol{\nu}\|_1}}
\boldsymbol{\nu}!\,,
\qquad
\forall\boldsymbol{\nu}\in\mathbb{N}_0^d\, .
\end{align}
\end{lemma}
\end{thmblock}

\paragraph{Refinement.} To build a good mesh of such hypercubes, we look at a single hypercube $H\subset D_T$ and we assume that it is chosen such that $f$ is analytic on $H$, in particular $H\cap S_F=\emptyset$. 
The error of the quadrature $\mathcal{Q}^H_N[f]$ on this hypercube depends on the distance $\mathrm{dist}(H,S)$ of the hypercube to the set of (complex) singularities $S$ of (the complex extension of) the integrand, the diameter of the hypercube $\text{diam}(H)$, and quadrature order $N$. We note that increasing the values of the propagator masses pushes the singularities from $S_E$ further into the complex domain, thereby improving the convergence of the algorithm. This should not come as a surprise: Feynman integrals may develop infrared singularities in the small-mass limit, and so large values of the masses render the integration more convergent.

 We have implemented an algorithm to partition the integration domain in a way  that optimizes the error. Mathematical details can be found in section~\ref{sec:hp-quadrature}. In a nutshell, our algorithm allows us to construct a mesh of differently-sized hypercubes $H$ on which we perform Gauss-Legendre quadrature. 
The meshes are generated recursively by accepting or rejecting hypercubes $H$ depending on the ratio
 \beq
 \max_{{i} \in [N_P]}\frac{\operatorname{diam}(H)}{\text{dist}(H,P_i)}\,.
 \eeq
 If a hypercube is rejected, then it is further partitioned in the next step, with a maximum number $M$ of such refinements. For each box we then choose the order of the Gauss-Legendre quadrature to be 
 \[
N= \lceil\alpha+(M-m)\beta\rceil\,,
 \]
 where $m$ is the number of subdivisions  (starting from the hypercube $D_T$) after which the hypercube is accepted. The two parameters $\alpha$ and $\beta$ are chosen in the end and control the error of the quadrature. 
 The closer we get to the singularity, the higher the refinement level becomes, and the smaller the cubes $H$ and the order of the quadrature rule are chosen.
 
 Let us make a few remarks. As explained in section~\ref{sec:history}, on a hypercube disjoint from the set $S$ of singular point, the quadrature error converges exponentially in the quadrature order $N$ for fixed $h$. Instead, refinement in $h$ yields only algebraic convergence for fixed $N$. Whenever the integrand is analytic, it is therefore generally more efficient to increase the quadrature order than to subdivide the cube. Accordingly, we use relatively large hypercubes with high quadrature order far from the singular set.
 Near the singular set, however, refinement is unavoidable for geometric reasons: As explained above, in order to cover as much of the integration domain as possible without any hypercube intersecting the singular set, the hypercubes must become progressively smaller as they approach it.  At the same time, the error estimate in eq.~\eqref{eq:GLerror} shows that the small volume of these hypercubes already suppresses their contribution to the total error. It is therefore unnecessary—and computationally inefficient—to assign them a very high quadrature order. We consequently decrease the quadrature order towards the singular set, while using higher orders on the larger hypercubes farther away.
 
 In the following we denote by $ \mathcal{H}_{(D_T,P,M, m, \alpha, \beta)}[f]$ the approximation of $\int_{\mathbb{R}^3} f(\mathbf{x})\,\rd \mathbf{x}$ using our $hp$-quadrature algorithm with the choice of parameters $(D_T,P,M, m, \alpha, \beta)$ introduced above.
 We will show in section~\ref{sec:hp-quadrature} that, if we prescribe a target error as 
 \begin{align}
 \Big|\int_{\mathbb{R}^3} f(\mathbf{x})\,\rd \mathbf{x}-\mathcal{H}_{(D_T,P,M, 0, \alpha, \beta)}[f]\Big|\leq 2 \sigma\,,
\end{align}
 then this error $2\sigma$ can be achieved for the following choice of parameters: 
  \begin{align*}
    \alpha &=1 \,,\quad \beta = 0.5\,,\\
            M &= \bigg\lceil\frac{\log_2(     (C(1+\sqrt{2})(4b\sqrt{3})^3)|P|)-\log_2(\sigma)+\log_2(3/2)3}{2}\bigg\rceil\,,
        \end{align*}
where in the cLTD representation specifically $|P|=N_P$ is the number of propagators, and $\lceil x \rceil$ is the closest integer greater than $x$. We also defined 
\begin{align}
C=
\sup_{\scriptstyle P_{\mathbf0}\left(\frac{\operatorname{dist}(H,S)}{\sqrt 3}\right)\oplus H}|f|\,.
\end{align}
Finally, our algorithm also depends on the parameter $b$ as introduced in eq.~\eqref{eq:defb}.

To summarise, we have developed an algorithm to produce a mesh of hypercubes $H$, and on each hypercube we may approximate the value of the integral by a Gauss-Legendre quadrature rules. 
We will show in particular that the algorithm converges with stretched-exponential convergence in the total number of function evaluations $E$ with
\begin{align}
 E = \mathcal{O}\Big(|P|M^{4}\Big)\,,
\end{align}
i.e., there are constants $a,c>0$ such that
        \begin{align}
        \left| \int_{D_T} f(\mathbf{x}) \,\rd\mathbf{x} -\mathcal{H}_{(D_T,P,M, 0, \alpha, \beta)}[f]\right|
        \leq
        a\exp\!\bigg(-c\frac{2}{\sqrt{3}}\bigg(\frac{E}{|P|}\bigg)^{1/4}\bigg)\,,
        \end{align}
        where $D_T$ as defined in eq.~\eqref{eq:defb}.
The residual domain $D_R$ (defined in eq.~\eqref{eq:defDR}) is integrated using very few quadrature nodes, resulting in significantly less evaluations than on the domain $D_T$. Hence the computational complexity is dominated by the integration of $D_T$. The convergence rate is unaffected by this integration.
        Note that the resulting stretched-exponential convergence is better than algebraic, but worse than exponential. It tells us that improving the approximation by a factor of two order requires only $\mathcal{O}(2^{4})$ more function evaluations. This allows us to obtain very precise results fairly cheaply.

\subsection{Details of the algorithm}
\label{sec:hp-quadrature}

In this section we present the mathematical details underlying our algorithm. In particular, we present rigorous mathematical proofs for the applicability of our $hp$-quadrature algorithm to one-loop integrals in $D=d+1=4$ space-time dimensions, and that it converges with stretched-exponential convergence. We focus our proofs on the case of one-loop integrals in Euclidean kinematics in the cLTD representation that are convergent in four dimensions, but whenever possible we present proofs that are valid in an arbitrary (integer) number of (spatial) dimensions $d$.

\subsubsection{Regularity of one-loop integrals}

We now show that one-loop integrals satisfy eq.~\eqref{eq:GLerroroverview}.
The key tool in the proof is the following general statement in approximation theory, which we will subsequently apply to our specific setting of Euclidean one-loop integrals. It is a direct consequence of ref.~\cite{boerm2010efficient}, theorem 4.16, and Hölder's inequality. 

\begin{thmblock}
 \begin{theorem} \label{thm:quad_error_scaling}
Let $f$ be analytic on ${H}$, i.e., let $f \in C^\infty({H})$, and let there be $C_f \in \mathbb{R}_{\geq 0}$, ${\gamma} \in \mathbb{R}_{>0}$ such that 
    \begin{align} \label{eqn:analytic_cond}
    \|\ \partial^{\boldsymbol{\nu}} f \|_{C({H})} \leq \frac{C_f}{\gamma^{\|\boldsymbol{\nu}\|_1}} \boldsymbol{\nu}!\,, \quad \forall \boldsymbol{\nu} \in \mathbb{N}^d_0\,.
    \end{align}
  The $N$-points Gauss-Legendre quadrature error on $H$, with $H$ as in \cref{eq:cube}, is bounded as
    \beq\label{eq:GLerror}
    \left| \int_H f(\mathbf{x})\, \rd\mathbf{x} - \mathcal{Q}^H_N[f]\right| \leq C h^d \left(1+\frac{h}{\gamma}\right)\left( \frac{h}{2\gamma + \sqrt{4h^2+\gamma^2}} \right)^{N}\,,
    \eeq
    where the constant $C$ depends on $C_f, N$ and $d$.
\end{theorem}
\end{thmblock}
We will also rely on the following result:
\begin{thmblock}
    \begin{theorem}\label{thm:general_holomorphic_derivative_estimate}
        Let $f\colon\mathbb{R}^d\to\mathbb{R}$. Assume that it has a  complex extension $\tilde{f}\colon\mathbb{C}^d\to\mathbb{C}$ which is holomorphic on $\mathbb{C}^d\setminus S$, where $S$ has Lebesgue measure zero in $\mathbb{C}^d$. Let $K \subset \mathbb{R}^d$ be compact, $K\cap S=\emptyset$. Then we have the following derivative estimate for every multi-index $\boldsymbol\alpha \in \mathbb{N}^d$:
\begin{equation}\label{eqn:anabounddist}
\|\partial^{\boldsymbol{\alpha}} f\|_{L^{\infty}(K)} \leq \frac{\boldsymbol{\alpha}!}{\gamma^{\|\boldsymbol{\alpha}\|_{1}}} \|f\|_{L^{\infty}(P_\mathbf{0}(\gamma) + K)}\,,
\qquad
\gamma=\frac{\operatorname{dist}(K,S)}{\sqrt{d}}\,,
\end{equation}
where $\operatorname{dist}(K,S)$ is distance defined in eq.~\eqref{eq:dist_def} and $P_\mathbf{a}(r)$ is a complex polydisc
\[
    P_\mathbf{a}(r) := \{(z_1,...,z_n) \in \mathbb{C}^n : \;|z_i-a_i| < r \quad \text{for all }i \in [d]\}
\]
of radius $r$ centered at $\mathbf{a} \in \mathbb{C}^d$. Here, `$+$' in $P_\mathbf{0}( \gamma) + K$ is understood as the Minkowski sum.
    \end{theorem}
\end{thmblock}
\begin{proof}
    This is a simple consequence of Cauchy's inequalities \cite{laurent2010holomorphic} for derivative bounds of holomorphic function and norm equivalence in finite dimensions.
\end{proof}

Since in our case the integration domain is real, $\gamma$ is strictly positive whenever $S\cap\mathbb{R}^d=\emptyset$, i.e., when the set of singularities is disjoint from the set of real-valued vectors. 
We note that $f$ in \cref{thm:general_holomorphic_derivative_estimate} will play the role of the cLTD integrand, and $S$ the role of the singularities of the complex extension of $f$. Combining \cref{thm:quad_error_scaling,thm:general_holomorphic_derivative_estimate} yields the following estimate on the quadrature error (cf.~eq.~\eqref{eq:GLerroroverview}).
\begin{thmblock}
\begin{corollary}\label{corr:quadrature_error_diam_dist}
Let $f$ and $S$ as in \cref{thm:general_holomorphic_derivative_estimate} and let $H$ be a cube as in \cref{eq:cube} with $\operatorname{dist}(H,S)>0$. Then the $N$-point Gauss-Legendre quadrature error on $H$ is bounded by
\begin{align*}
&\left| \int_H f(\mathbf{x}) \,\rd\mathbf{x} - \mathcal{Q}^H_N[f]\right|\\
&\qquad\leq C \operatorname{diam}(H)^d \left(1+\frac{\operatorname{diam}(H)}{\operatorname{dist}(H,S)}\right)\left( \frac{\operatorname{diam}(H)}{2\operatorname{dist}(H,S) + \sqrt{4\operatorname{diam}(H)^2+\operatorname{dist}(H,S)^2}} \right)^{N},
\end{align*}
where $\operatorname{diam}(H)$ is the diameter, for any hypercube in $\mathbb{R}^d$ which has a positive distance to $S$.
\end{corollary}
\end{thmblock}

Next, it will be important to provide numerically accessible lower bounds for $\operatorname{dist}(H,S)$.  
We decompose the set of singularities $S = S_E \cup S_F$  as  in eqs.~\eqref{eq:SESF} and~\eqref{eq:SESF2} and define $P$ as in eq.~\eqref{eqn:sing_point_set}.
We assume that $S_E \cap \mathbb{R}^d = \emptyset$,\footnote{This is inspired by the setup for one-loop integrals in the Euclidean region where the relevant singularities are ellipsoids $S_E$ (which do not intersect the real integration domain, however have singularities in the complex extension) and focal points $S_F$.} implying $\operatorname{dist}(H,S_E)>0$ for all hypercubes $H$. Thus, in view of $\operatorname{dist}(H,S)=\min\{\operatorname{dist}(H,S_F),\operatorname{dist}(H,S_E)\}$, we only need to guarantee $\operatorname{dist}(H,S_F) > 0$, which in turn implies $\operatorname{dist}(H,S)>0$.

\bigskip
\begin{thmblock}
    \begin{corollary}\label{corr:gamma_dist_norm}
        Let $S_F \subset \mathbb{C}^d$ be as in \cref{eq:SESF} and $P \equiv S_F \cap \mathbb{R}^d$ be the point set defined in \cref{eqn:sing_point_set}. Then the following inequality holds:
        \[
        \operatorname{dist}(H,S_F) \geq \frac{\operatorname{dist}(H,P)}{\sqrt{2}}\,.
        \]
    \end{corollary}
\end{thmblock}
\begin{proof}
    Let $\mathbf{z}\in S_{F,i}$, $i\in[N_P]$, and $\mathbf{x}\in H$ arbitrary but fixed. Then $\mathbf{z}\in S_{F,i}$ and $m_i>0$ imply $(\mathbf{z}-\boldsymbol{\alpha}_i)\cdot (\mathbf{z}-\boldsymbol{\alpha}_i)\in\mathbb{R}_{<0}$  and thus $\|\operatorname{Re}(\mathbf{z})-\boldsymbol{\alpha}_i\|_2 < \|\operatorname{Im}(\mathbf{z})\|_2$. This yields
    \begin{align*}
        \operatorname{dist}(\mathbf{x},\mathbf{z})^2
        &=
        (\mathbf{x}-\mathbf{z})\cdot\overline{(\mathbf{x}-\mathbf{z})}\\
        &=
        \|\mathbf{x}-\operatorname{Re}(\mathbf{z})\|_2^2+\|\operatorname{Im}(\mathbf{z})\|_2^2\\
        &\geq
        \|\mathbf{x}-\operatorname{Re}(\mathbf{z})\|_2^2+\|\operatorname{Re}(\mathbf{z})-\boldsymbol{\alpha}_i\|_2^2
        \geq
        \frac{\|\mathbf{x}-\boldsymbol{\alpha}_i\|_2^2}{2}\,.
    \end{align*}
    This yields the assertion.
\end{proof}
Thus, verifying that $\operatorname{dist}(H,P) > 0$ is enough to conclude that $\operatorname{dist}(H,S) > 0$. 
This corollary informs us that, analogously to corollary \ref{corr:quadrature_error_diam_dist}, the quadrature error scaling for our finite, Euclidean, one-loop setup only depends on the size of the hypercube $H$ and its distance to the closest point in the point set $P$ when $\sqrt{2}\operatorname{dist}(H,S_F) \leq \operatorname{dist}(H,S_E)$. More precisely, we have
\begin{align} \label{eq:quadrature_error_point_dist}
\left| \int_H f(\mathbf{x}) \,\rd\mathbf{x} - \mathcal{Q}^H_N[f]\right| \leq &C h^d \left(1+\sqrt{2}\frac{\operatorname{diam}(H)}{\operatorname{dist}(H,P)}\right)\\
&\,\times\left( \frac{\sqrt{2}\operatorname{diam}(H)}{2\operatorname{dist}(H,P) + \sqrt{8\operatorname{diam}(H)^2+\operatorname{dist}(H,P)^2}} \right)^{N}\,.\nonumber
\end{align}

\subsubsection{$hp$-quadrature}
We split the integration domain as in eqs.~\eqref{eq:defDR} and~\eqref{eq:defb} into a \textit{total-field} $D_T$ and \textit{residual} $D_{R}$ such that 
\begin{equation}\label{eq:residualsplitting}
    S \cap \mathbb{R}^d\equiv P\subset D_T := [-b,b]^d\,, \qquad D_R:=\mathbb{R}^d \backslash D_T\,.
\end{equation}
To avoid technicalities, we assume $b\geq 1$. Should this be unreasonable, we can rescale the problem to more reasonable scales.

In the following we address the integration of the domain $D_T$ containing the set $S$ of all singular points. For the numerical integration on $D_R$ up to desired precision we refer to appendix~\ref{app:choiceofb}. For the numerical integration on $D_T$, we observe from eq.~\eqref{eq:quadrature_error_point_dist} that achieving a good error decay requires the integration domain neither to contain nor to touch any point in $P$. Even when the domain $D_T$ is partitioned into smaller domains, this condition cannot be satisfied unconditionally, since some elements will inevitably contain or border such points. Our goal is to construct a partition in which these problematic elements are sufficiently small that their contributions are negligible and can be safely ignored.

Equation (\ref{eq:quadrature_error_point_dist}) further indicates that the ratio between the distance to $P$ and the diameter of an element $H$ plays an important role in the error scaling. We adopt a simple ratio test in our integration algorithm to determine whether partition elements should be accepted or rejected.
More precisely, we perform a divide-and-conquer strategy based on the recursive dyadic partition of $D_T$. We perform the partition on each newly obtained element until all of them satisfy the admissibility condition

\begin{equation}\label{eq:admissibility}
\mu_P(H) := \max_{{i} \in [N_P]}\frac{\operatorname{diam}(H)}{\text{dist}(H,P_i)} \leq 1\,,
\end{equation}
implying ${\operatorname{dist}(H,S)} \geq \gamma'' >0$, where $H$ is a hypercube with a side of size $h$. We say that the element $H$ is admissible if and only if it satisfies the condition function $\mu_P$.
This leads to the recursive partition-integration algorithm as given below:

\begin{algorithm}[H]
\caption{\label{alg:cap}$hp$-quadrature on \textit{total-field}}
\begin{algorithmic}
\State \textbf{Input:} Box $H$, focal point set $P$, max. refinements $M$, refinement level $m$, integrand $f$
\State \textbf{Output:} Approximation of \cref{LTDTriangleCausal} on integration domain $H$
\State \textbf{Procedure:} $\mathcal{H}_{(H,P,M, m, \alpha, \beta)}[f]$
\If{$H$ satisfies admissibility $\mu_P(H)$}
\State \Return $\mathcal{Q}_{\lceil\alpha+(M-m)\beta\rceil}^H[f]$
\ElsIf{$m \leq M$} \Comment{sufficiently small elements are discarded}
\State Generate dyadic partition $\{H_i\}_{i=1}^{2^d}$ of $H$
\State \Return $\sum_{i=1}^{2^d}$$\mathcal{H}_{(H_i,P, M,m+1, \alpha, \beta)}[f]$
\EndIf
\State \Return 0
\end{algorithmic}
\end{algorithm}

The algorithm partitions the domain into a set of subdomains on which $f$ is then integrated or set to zero. For further analysis,
we introduce the sets
\begin{align*}
    \mathfrak{N}&=\{\text{inadmissible subdomains for which \cref{alg:cap} returns 0}\}\,,\\
    \mathfrak{F}_\ell&=\{\text{admissible subdomains which \cref{alg:cap} integrates with $m=\ell$}\}\,,\qquad \ell=0,\ldots,M\,,
\end{align*}
and
\begin{equation}
\mathfrak{F}=\bigcup_{m=0}^M\mathfrak{F}_m\,.
\end{equation}
As we will see, the asymptotic cardinality of $\mathfrak{F}_m$ and the cardinality of $\mathfrak{N}$ are important for our considerations.

\begin{thmblock}
    \begin{lemma}\label{lemma:total_level_elem_count}
        There is $m_0\in\mathbb{N}$ such that $|\mathfrak{F}_m|=\mathcal{O}(|P|2^dd^{d/2})$ for all $m\geq m_0$, with constants depending only on geometric quantities, and not on $M,f,\alpha$, and $\beta$.
    \end{lemma}
\end{thmblock}
\begin{proof}
    Let $\mathcal{C}_m$ the set of hypercubes after $m$ dyadic partition steps of $D_T$ and $\sqrt{d} h_m=\operatorname{diam}(\square)$, $\square\in\mathcal{C}_m$, the constant diameter of the hypercubes therein. Then
    \[
    \mathfrak{F}_m
    \subset
    \bigcup_{\boldsymbol{\alpha}\in P}\Big\{\square\in\mathcal{C}_m\colon \sqrt{d} h_m<\operatorname{dist}(\square,\boldsymbol{\alpha})\leq 2\sqrt{d} h_m\Big\}\,.
    \]
    Thus,
    \begin{align*}
    |\mathfrak{F}_m|
    &\leq
    \sum_{\boldsymbol{\alpha}\in P}\Big|\Big\{\square\in\mathcal{C}_m\colon \sqrt{d} h_m<\operatorname{dist}(\square,\boldsymbol{\alpha})\leq 2\sqrt{d} h_m\Big\}\Big|\\
    &\leq
    |P|\max_{\boldsymbol{\alpha}\in P}\Big|\Big\{\square\in\mathcal{C}_m\colon \sqrt{d} h_m<\operatorname{dist}(\square,\boldsymbol{\alpha})\leq 2\sqrt{d} h_m\Big\}\Big|\,.
    \end{align*}
    We estimate the maximum by finding an upper bound on how many hypercubes in $\mathcal{C}_m$ intersect $B_{2\sqrt{d} h_m}(\boldsymbol{0})\setminus B_{\sqrt{d} h_m}(\boldsymbol{0})$, where $B_{r}(\boldsymbol{0})$ is the closed ball of radius $r>0$. Using a volume argument we obtain that for $m$ sufficiently large it holds
    \begin{align*}
    \max_{\boldsymbol{\alpha}\in P}\Big|\Big\{\square\in\mathcal{C}_m\colon \sqrt{d} h_m<\operatorname{dist}(\square,\boldsymbol{\alpha})\leq 2\sqrt{d} h_m\Big\}\Big|
    &\leq
    \Big|\Big\{\square\in\mathcal{C}_m\colon \sqrt{d} h_m<\operatorname{dist}(\square,\boldsymbol{0})\leq 2\sqrt{d} h_m\Big\}\Big|\\
    &\leq
    \frac{|B_{3\sqrt{d} h_m}(\boldsymbol{0})|-|B_{\sqrt{d} h_m}(\boldsymbol{0})|}{h_m^d}\\
    &=
    \frac{(3\sqrt{d} h_m-\sqrt{d} h_m)^d|B_{1}(\boldsymbol{0})|}{h_m^d}\\
    &=
    2^dd^{d/2}|B_{1}(\boldsymbol{0})|\,.
    \end{align*}
    This yields the assertion.
\end{proof}
\subsubsection{$hp$-quadrature error}

For the error analysis of of the $hp$-quadrature, it is convenient to split the integral over $D_T$ as
\[
\int_{D_T}f(\mathbf{x})\,\rd\mathbf{x}
=
\underbrace{
\int_{D_T\setminus R_U}f(\mathbf{x})\,\rd\mathbf{x}
}_{=(\spadesuit)}
+
\underbrace{\int_{R_U}f(\mathbf{x})\,\rd\mathbf{x}}_{=(\clubsuit)}\,,
\]
where
\[
R_U
=
\bigcup_{\square\in\mathfrak{N}}\square\,.
\]
In \cref{alg:cap}, $(\clubsuit)$ is approximated by zero, whereas $(\spadesuit)$ is approximated by a sophisticated combination of tensorized Gauss-Legendre rules. The error analysis is thus split into two parts. First, the truncation of error of setting  $(\clubsuit)$ to zero, and, second, the approximation of $(\spadesuit)$ .

\paragraph{Truncation error.}
The only way to control the truncation error $(\clubsuit)$ is to adapt the size of $R_U$, which is controlled by the number of maximum refinements $M$. In the following lemma, we give a lower bound on $M$ to achieve a certain truncation error.
\begin{thmblock}
    \begin{lemma}\label{lemma:minimum_refinment_levels}
        For the $R_U\subset \mathbb{R}^d$ and the integrand $f$ to satisfy
        \begin{align}\label{eqn:ru_sigma_error}
        \left| \int_{R_U}f(\mathbf{x})\,\rd\mathbf{x} \right| \leq \sigma\,,
        \end{align}
        where $\sigma \in \mathbb{R}_{>0}$, we need to ensure that we have at least
        \begin{align}\label{eqn:M_lower_bound}
        M \geq \frac{\log_2(\hat{C}|P|)-\log_2(\sigma)+\log_2(3/2)d}{d-1}
        \end{align}
        levels of refinement in \cref{alg:cap}, where
        \begin{align}\label{eqn:c_hat_constant}
        \hat{C}:= C \left(1+\sqrt{2} \right)\Big(4b\sqrt{d}\Big)^d.
        \end{align}
    \end{lemma}
\end{thmblock}
\begin{proof}
Looking at the integrand $f$ from \cref{LTDTriangleCausal}, it is easy to see that there is $r\in\mathbb{R}_{>0}$ such that for all $\boldsymbol{\alpha}\in P$ it holds
\[
|f(\mathbf{x})| \leq \frac{C}{\|\mathbf{x}-\boldsymbol{\alpha}\|_2},\qquad\text{for all}~\mathbf{x}\in B_r(\boldsymbol{\alpha})\,.
\]
By a similar argumentation as in the proof of \cref{lemma:total_level_elem_count} we obtain
\begin{align*}
    \left| \int_{R_U} f(\mathbf{x}) \,\rd \mathbf{x} \right|
    &\leq \int_{R_U} |f(\mathbf{x})| \,\rd\mathbf{x} \\
    &\leq \sum_{\boldsymbol{\alpha}\in P}\int_{B_{3\sqrt{d}h_M}(\mathbf{0})} |f(\mathbf{x})| \,\rd\mathbf{x}\\
    &\leq C|P| \int_{B_{3\sqrt{d}h_M}(\mathbf{0})}\|\mathbf{x}\|^{-1}_2 \,\rd \mathbf{x} \\
    &\leq C|P|3^{d-1}h_M^{d-1}\frac{d^{(d+1)/2}}{d-1}|B_1(\boldsymbol{0})|\,.
\end{align*}
Inserting \cref{eqn:c_hat_constant} yields
\begin{align*}
    \left| \int_{R_U} f(\mathbf{x}) \,\rd \mathbf{x} \right| \leq \hat{C} |P|2^{-((d-1)M+\log_2(2/3)d)}\,,
\end{align*}
as $b \geq 1$ for $d > 1$. From this we know that if we want to ensure eq.~\eqref{eqn:ru_sigma_error},
it is sufficient to lower bound the minimum number of refinement levels with eq.~\eqref{eqn:M_lower_bound}.
\end{proof}

\paragraph{Quadrature error.}
Estimating the approximation error for $(\spadesuit)$ using \cref{alg:cap} is given by the sum of errors from the tensorized quadrature rules. We will show that by controlling the quadrature order with constants $\alpha$ and $\beta$ we can obtain any desired error $\sigma$.
\begin{thmblock}
    \begin{lemma} \label{lemma:sufficient_order_of_quadrature}
    Let $\beta\in\mathbb{N}$ and $\sigma\in\mathbb{R}_{>0}$. Let $\kappa := 5/\sqrt{2}$ and $\alpha\in\mathbb{N}_0$ satisfy
    \begin{equation}\label{eq:alphabound}
    \alpha\kappa
    \geq
    \log_2(\hat{C}|P|)-\log_2(\sigma)-M\min(d,\beta\kappa)-\log_2(1-2^{-|d-\beta\kappa|})\,,
    \end{equation}
    where $\hat{C}$ is the constant from \cref{eqn:c_hat_constant}. Then it holds
    \[
    \left| \int_{D_T \backslash R_U} f(\mathbf{x}) \,\rd\mathbf{x} - \mathcal{H}_{(D_T,P,M,0,\alpha ,\beta)}[f]\right| \leq \sigma\,.
    \]
    \end{lemma}
\end{thmblock}
\begin{proof}
    For $\square\in\mathfrak{F}_m$ combining \cref{eq:admissibility} with \cref{eq:quadrature_error_point_dist} yields
     \begin{align*}
    \left| \int_{\square} f(\mathbf{x}) \,\rd\mathbf{x} - \mathcal{Q}^{\square}_{\alpha+(M-m)\beta}[f]\right| &  \leq C \left(1+\sqrt{2} \right)|\square| 2^{-(\alpha+(M-m)\beta)\kappa} \\
    & \leq C \left(1+\sqrt{2} \right)(2b)^d 2^{-\left(dm+(\alpha+(M-m)\beta)\kappa \right)}\,.
    \end{align*}
    The estimate is uniform for all $\square\in\mathfrak{F}_m$. Using \cref{lemma:total_level_elem_count} we obtain
    \begin{align*}
    &\left| \int_{D_T \backslash R_U} f(\mathbf{x}) \,\rd\mathbf{x} - \mathcal{H}_{(D_T,P,M,0,\alpha ,\beta)}[f]\right|\\
    &\qquad\leq\sum_{m=0}^M\sum_{\square\in\mathfrak{F}_m}\left| \int_{\square} f(\mathbf{x}) \,\rd\mathbf{x} - \mathcal{Q}^{\square}_{\alpha+(M-m)\beta}[f]\right|\\
    &\qquad\leq C \left(1+\sqrt{2} \right)(2b)^d\sum_{m=0}^M\sum_{\square\in\mathfrak{F}_m} 2^{-\left(dm+(\alpha+(M-m)\beta)\kappa \right)}\\
    &\qquad \leq \hat{C} |P| \sum_{m=0}^M 2^{-\left(dm+{\alpha}\kappa+(M-m){\beta}\kappa \right)}\\
    &\qquad \leq \hat{C} |P| 2^{-({\alpha\kappa+M\min(d,\beta\kappa)})} \sum_{m=0}^\infty 2^{-m\left|d-{\beta}\kappa \right|}\\
    &\qquad \leq  \hat{C} |P| 2^{-({\alpha\kappa+M\min(d,\beta\kappa)})} \frac{1}{1-2^{-\left|d-\beta\kappa \right|}}\,.
    \end{align*}
    We can note that due to $\kappa$ being irrational, we obtain $d \neq \beta \kappa$, making the latter expression well-defined. This yields the assertion.
\end{proof}
\begin{remark}\label{rem:alphabetafloat}
Although $\alpha$ and $\beta$ have been defined as integers, in practice it is easier and computationally cheaper to calculate them as real numbers and round up the expected quadrature order to a suitable integer.
\end{remark}

\paragraph{Overall choice of parameters for $hp$-quadrature \cref{alg:cap}.}
To achieve an overall error of the $hp$-quadrature algorithm of $2\sigma$, the parameters $\alpha,\beta$, and $M$ need to be chosen such that \cref{lemma:minimum_refinment_levels} and \cref{lemma:sufficient_order_of_quadrature} hold simultaneously. Combining these results yields the following theorem, indicating in particular that $\alpha$ and $\beta$ can be chosen independently of $M$.
\begin{thmblock}
    \begin{theorem}\label{thm:chooseparameters}
        Let $\sigma\in (0,1]$. Choose
        \begin{align*}
            M &= \bigg\lceil\frac{\log_2(\hat{C}|P|)-\log_2(\sigma)+\log_2(3/2)d}{d-1}\bigg\rceil,\\
            \beta &= \bigg\lceil\frac{d-1}{\kappa}\bigg\rceil,\\
            \alpha &= \bigg\lceil-\frac{\log_2\big(1-2^{-|d-\beta\kappa|}\big)}{\kappa}\bigg\rceil,
        \end{align*}
        in \cref{alg:cap}, with $\hat{C}$ as in \cref{eqn:c_hat_constant}. Then it holds
        \[
        \left| \int_{D_T} f(\mathbf{x}) \,\rd\mathbf{x} - \mathcal{H}_{(D_T,P,M,0,\alpha ,\beta)}[f]\right|
        \leq
        2\sigma.
        \]
    \end{theorem}
\end{thmblock}
\begin{proof}
Strengthen the bound \cref{eq:alphabound} by inserting the lower bound on $M$ from \cref{eqn:M_lower_bound} yields the desired result.
\end{proof}

 \begin{remark} 
$\beta$ in theorem~\ref{thm:chooseparameters} is defined as an integer. In practice, this can lead to unnecessarily high quadrature orders, and we choose a lower, possibly non-integer, value.
 \end{remark}

\paragraph{Stretched-exponential convergence.}

In order to estimate the runtime complexity of the \cref{alg:cap}, we need to acknowledge that the function evaluations of $f$ are the most expensive part of this process. It is easy to show that the runtime complexity is algebraic with respect to number of levels and exponential with respect to the dimension.
\begin{thmblock}
    \begin{theorem} \label{thm:eval_complexity}
    Let $E$ be total number evaluations of function f during the runtime of \cref{alg:cap}. This number of evaluations $E$ satisfies
    \[
    E = \mathcal{O}\Big(|P|2^dd^{d/2}M^{d+1}\Big),
    \]
    where the constants depend on $\alpha$ and $\beta$.
    \end{theorem}
\end{thmblock}
\begin{proof}
    Let $m'=\min\{m\in\mathbb{N}\colon\mathfrak{F}_m\neq\emptyset\}$. This number is independent of $M$. Using \cref{lemma:total_level_elem_count} we obtain
    \begin{align*}
    E &= \sum_{m=0}^M\sum_{\square\in\mathfrak{F}_m}(\alpha+(M-m)\beta)^d\\
    &= \sum_{m=m'}^M\sum_{\square\in\mathfrak{F}_m}(\alpha+(M-m)\beta)^d\\
    &= |\mathfrak{F}_m|(\sum_{m=m'}^M\alpha+(M-m)\beta)^d\\
    &= \mathcal{O}\bigg(|P|2^dd^{d/2}\sum_{m=m'}^M (\alpha+(M-m)\beta)^d\bigg)\\
    &= \mathcal{O}\bigg(|P|2^dd^{d/2}\sum_{m=0}^M (\alpha+m\beta)^d\bigg) \\
    &= \mathcal{O}\bigg(|P|2^dd^{d/2}(M+1)(\alpha + M\beta)^d\bigg)\,,
    \end{align*}
    which yields the assertion.
\end{proof}
This yields the observation that the quadrature error of \cref{alg:cap} is streched-exponential with respect to the total number of function evaluations.
\begin{thmblock}
    \begin{corollary}
        Under the assumptions of \cref{thm:chooseparameters}, the $hp$-quadrature \cref{alg:cap} converges stretched-exponentially in the number of function evaluations $E$, i.e., there are constants $a,b>0$ such that
        \[
        \left| \int_{D_T} f(\mathbf{x}) d\mathbf{x} - \mathcal{H}_{(D_T,P,M,0,\alpha ,\beta)}[f]\right|
        \leq
        a\exp\bigg(-b\frac{(d-1)}{\sqrt{d}}\bigg(\frac{E}{|P|}\bigg)^{1/(d+1)}\bigg)\,.
        \]
    \end{corollary}
\end{thmblock}
\begin{proof}
    From \cref{thm:chooseparameters} we observe that choosing $M(d-1)\sim -\log_2(\sigma)$ is sufficient to obtain an approximation error $2^{-M(d-1)}\sim\sigma>0$. From \cref{thm:eval_complexity} we obtain that this requires
    \[
    E
    =
    \mathcal{O}\Big(|P|2^{d+1}d^{(d+1)/2}M^{d+1}\Big)
    \]
    function evaluations, implying
    \[
    M
    \geq
    CE^{1/(d+1)}|P|^{-1/(d+1)}d^{-1/2}.
    \]
    This yields the assertion.
\end{proof}

 \begin{remark}     
 
 \begin{enumerate} 
\item The worsening of the rate with increasing dimension is to be expected as the dimensionality affects the maximum number of admissible domains exponentially, as shown in \cref{thm:eval_complexity}.
\item Since wast majority of the evaluations are spent in the asymptotic convergence regimes in the vicinity of $P$, the worsening of the error as reflected in the estimate is convincing.
\end{enumerate}
\end{remark}

\section{Examples and benchmarks}
\label{sec.examples}

As a proof of concept, we have implemented our $hp$-quadrature algorithm into a {\tt C++} code to evaluate finite one-loop integrals in the Euclidean region. In the previous section, we have seen that our algorithm converges with a stretched-exponential convergence with the number of evaluations. In order to gauge the performance of our code, we provide in this section benchmark evaluations of the triangle, box and pentagon integrals depicted in figure~\ref{fig:topologies} at selected (Euclidean) kinematic points, and we compare our results to a numerical evaluation of the same cLTD representation using other numerical integration techniques and analytic results.
The code is available from \begin{center}
\href{https://github.com/BKova/feynman-one-loop-hp-quadrature}{{\tt https://github.com/BKova/feynman-one-loop-hp-quadrature}}. 
\end{center}

\begin{figure}[ht]
    \centering

    \begin{minipage}[t]{0.31\textwidth}
        \vspace{0pt}
        \centering
        \resizebox{\linewidth}{!}{%
        \begin{tikzpicture}[
            thick,
            line cap=round,
            line join=round,
            extmom/.style={
                postaction={decorate},
                decoration={
                    markings,
                    mark=at position 0.55 with {\arrow{>}}
                }
            }
        ]

        \path[use as bounding box]
            (-2.9,-2.95) rectangle (2.9,2.95);

        \coordinate (A) at ( 0.000,  1.400);
        \coordinate (B) at (-1.617, -1.400);
        \coordinate (C) at ( 1.617, -1.400);

        \coordinate (P1) at ( 0.000,  2.550);
        \coordinate (P2) at ( 2.500, -2.550);
        \coordinate (P3) at (-2.500, -2.550);

        \draw (A) -- node[above right, pos=0.52] {$m_1$} (C);
        \draw (C) -- node[below, pos=0.50] {$m_2$} (B);
        \draw (B) -- node[above left, pos=0.48] {$m_3$} (A);

        \draw[extmom] (P1) -- (A)
            node[pos=0, above] {$p_1$};

        \draw[extmom] (P2) -- (C)
            node[pos=0, below right] {$p_2$};

        \draw[extmom] (P3) -- (B)
            node[pos=0, below left] {$p_3$};

        \end{tikzpicture}%
        }
    \end{minipage}
    \hfill
    \begin{minipage}[t]{0.31\textwidth}
        \vspace{0pt}
        \centering
        \resizebox{\linewidth}{!}{%
        \begin{tikzpicture}[
            thick,
            line cap=round,
            line join=round,
            extmom/.style={
                postaction={decorate},
                decoration={
                    markings,
                    mark=at position 0.55 with {\arrow{>}}
                }
            }
        ]

        \path[use as bounding box]
            (-2.9,-2.95) rectangle (2.9,2.95);

        \coordinate (A) at (-1.400,  1.400);
        \coordinate (B) at ( 1.400,  1.400);
        \coordinate (C) at ( 1.400, -1.400);
        \coordinate (D) at (-1.400, -1.400);

        \coordinate (P1) at (-2.150,  2.550);
        \coordinate (P2) at ( 2.150,  2.550);
        \coordinate (P3) at ( 2.150, -2.550);
        \coordinate (P4) at (-2.150, -2.550);

        \draw (A) -- node[above, pos=0.5] {$m_1$} (B);
        \draw (B) -- node[right, pos=0.5] {$m_2$} (C);
        \draw (D) -- node[below, pos=0.5] {$m_3$} (C);
        \draw (A) -- node[left, pos=0.5] {$m_4$} (D);

        \draw[extmom] (P1) -- (A)
            node[pos=0, above left] {$p_1$};

        \draw[extmom] (P2) -- (B)
            node[pos=0, above right] {$p_2$};

        \draw[extmom] (P3) -- (C)
            node[pos=0, below right] {$p_3$};

        \draw[extmom] (P4) -- (D)
            node[pos=0, below left] {$p_4$};

        \end{tikzpicture}%
        }
    \end{minipage}
    \hfill
    \begin{minipage}[t]{0.31\textwidth}
        \vspace{0pt}
        \centering
        \resizebox{\linewidth}{!}{%
        \begin{tikzpicture}[
            thick,
            line cap=round,
            line join=round,
            extmom/.style={
                postaction={decorate},
                decoration={
                    markings,
                    mark=at position 0.55 with {\arrow{>}}
                }
            }
        ]

        \path[use as bounding box]
            (-2.9,-2.95) rectangle (2.9,2.95);

        \coordinate (A) at (-0.910,  1.400);
        \coordinate (B) at ( 0.910,  1.400);
        \coordinate (C) at ( 1.472, -0.330);
        \coordinate (D) at ( 0.000, -1.400);
        \coordinate (E) at (-1.472, -0.330);

        \coordinate (P1) at (-1.650,  2.550);
        \coordinate (P2) at ( 1.650,  2.550);
        \coordinate (P3) at ( 2.550, -0.600);
        \coordinate (P4) at ( 0.000, -2.550);
        \coordinate (P5) at (-2.550, -0.600);

        \draw (A) -- node[above] {$m_1$} (B);
        \draw (B) -- node[right] {$m_2$} (C);
        \draw (C) -- node[below right] {$m_3$} (D);
        \draw (D) -- node[below left] {$m_4$} (E);
        \draw (E) -- node[left] {$m_5$} (A);

        \draw[extmom] (P1) -- (A)
            node[pos=0, above left] {$p_1$};

        \draw[extmom] (P2) -- (B)
            node[pos=0, above right] {$p_2$};

        \draw[extmom] (P3) -- (C)
            node[pos=0, right] {$p_3$};

        \draw[extmom] (P4) -- (D)
            node[pos=0, below] {$p_4$};

        \draw[extmom] (P5) -- (E)
            node[pos=0, left] {$p_5$};

        \end{tikzpicture}%
        }
    \end{minipage}

    \caption{Triangle, box, and pentagon integrals implemented into our code. 
    }
    \label{fig:topologies}
\end{figure}

\subsection{One-loop integrals in Euclidean kinematics}\label{sec:finiteoneloop}
 We focus on the triangle, box and pentagon integrals from figure~\ref{fig:topologies}, and we choose the external kinematics to lie in four space-time dimensions. The kinematic points are given in tables~\ref{tab:triangle-points},~\ref{tab:box-points} and~\ref{tab:pentagon-points} for the triangle, box and pentagon integrals, respectively.
 Note that some of these kinematic configurations were also computed in~refs.~\cite{Buchta:2015wna,Capatti:2019ypt}.
We compare the performance of our $hp$-quadrature algorithm to results obtained in \texttt{Python} with the \texttt{Vegas} implementation from the \texttt{Cuba} library~\cite{Hahn:2004fe}.  
Additionally, whenever available, we compare the numerical results to the analytic expressions provided by the  \texttt{LoopTools} library~\cite{Hahn:1998yk}.\footnote{To increase the numerical accuracy, we use the precision extension of the library, enabling us to reach quadruple precision.} The cLTD representations of all integrands were derived using the package of ref.~\cite{Capatti:2020ytd}.
For the $hp$-quadrature algorithm,  $M$ is the maximum number of refinements as chosen in theorem~\ref{thm:chooseparameters} and $\alpha$ and $\beta$ are the parameters that govern the quadrature order as used in the algorithm.  For our numerical results we choose $\beta=0.5$, using the expressions from theorem~\ref{thm:chooseparameters}.  The parameter $b$ from eq.~\eqref{eq:defb} is chosen as explained in appendix~\ref{app:choiceofb}.

We evaluate the integrand several million times during the
\texttt{VEGAS} integrations. Although the exported file is read
and parsed only once, a direct {\tt Python} implementation still evaluates the
integrand through the {\tt Python} interpreter separately for every Monte Carlo
sample. For sufficiently large sample counts, this leads
to prohibitively long runtimes.
To avoid this bottleneck, a {\tt C} backend was implemented by
\texttt{GPT-5.6 Sol}. The backend translates the exported integrand and its
helper definitions into {\tt C} source code and invokes the {\tt C} compiler once. The
result is a native machine-code function of the final integrand.
\texttt{VEGAS} passes sample points to this function in batches. Within each
batch, the compiled function evaluates all sample points in a single {\tt C} loop
and returns the resulting array only after the batch is complete. Thus, there
is no return to the {\tt Python} interpreter between individual sample points in the
same batch.

\begin{table}[!t]
  \centering
  \small
  \renewcommand{\arraystretch}{1.15}
  \begin{tabular}{@{}ccl@{}}
    \toprule
    Configuration & Quantity & Value \\
    \midrule

    \multirow{3}{*}{\(P_1^\triangle\)}
      & \(p_1\) & \((1,3,4,5)\) \\
      & \(p_2\) & \((1,6,7,8)\) \\
      & \(m\)   & \((0,0,0)\) \\
    \addlinespace

    \multirow{3}{*}{\(P_2^\triangle\)}
      & \(p_1\) & \((5.23923,-4.18858,0.74966,-3.05669)\) \\
      & \(p_2\) & \((6.99881,-2.93659,5.03338,3.87619)\) \\
      & \(m\)   & \((7.73358,7.73358,7.73358)\) \\
    \addlinespace

    \multirow{3}{*}{\(P_3^\triangle\)}
      & \(p_1\) & \((13.4225,58.7948,-73.1186,-91.9502)\) \\
      & \(p_2\) & \((81.6593,-68.5217,8.75578,-95.0535)\) \\
      & \(m\)   & \((49.9745,86.9249,80.2257)\) \\
    \addlinespace

    \multirow{3}{*}{\(P_4^\triangle\)}
      & \(p_1\) & \((-2,11,5,-3)\) \\
      & \(p_2\) & \((10,15,14,12)\) \\
      & \(m\)   & \((11,8,2)\) \\
    \addlinespace
    \bottomrule
  \end{tabular}
  \caption{The considered kinematic points for the triangle integrals in figure~\ref{fig:topologies}. All these points lie in the Euclidean region.  
  Four-momenta are given in the convention
  \(p_i=(p_i^{(0)},p_i^{(1)},p_i^{(2)},p_i^{(3)})\) and \(m=(m_1,m_2,m_3)\) denotes the
  vector of internal masses. The  numerical results for the $P_i^\triangle$ are shown in table~\ref{tab:triangle-leading-results}.}
    \label{tab:triangle-points}
\end{table}

\begin{table}[!th]
  \centering
  \small
  \renewcommand{\arraystretch}{1.15}
  \begin{tabular}{@{}ccl@{}}
    \toprule
    Configuration & Quantity & Value \\
    \midrule

    \multirow{4}{*}{\(P_1^\Box\)}
      & \(p_1\) & \((31.5487,-322.403,300.53,-385.58)\) \\
      & \(p_2\) & \((103.904,202.01,-451.278,-435.128)\) \\
      & \(p_3\) & \((294.767,252.89,447.092,311.716)\) \\
      & \(m\)   & \((4.68481,4.68481,4.68481,4.68481)\) \\
    \addlinespace

    \multirow{4}{*}{\(P_2^\Box\)}
      & \(p_1\) & \((50.8543,-55.7461,11.6999,94.9259)\) \\
      & \(p_2\) & \((0.69914,67.1926,-5.78627,91.5278)\) \\
      & \(p_3\) & \((52.3577,76.3226,43.8222,13.0587)\) \\
      & \(m\)   & \((54.2965,53.5406,55.9681,51.7444)\) \\
    \addlinespace

    \multirow{4}{*}{\(P_3^\Box\)}
      & \(p_1\) & \(\left(-2,-1,-1,-\sqrt{2}\right)\) \\
      & \(p_2\) & \(\left(-2,-1,-1,-\sqrt{2}\right)\) \\
      & \(p_3\) & \(\left(-2,-1,-1,-\sqrt{2}\right)\) \\
      & \(m\)   & \((0.25,0.5,1,0)\) \\
    \addlinespace
    \midrule
    \addlinespace

    \multirow{4}{*}{\(P_4^\Box\)}
      & \(p_1\) &
        \(\left(-\sqrt{0.6},
        \frac{0.5}{\sqrt{0.6}},
        0,
        \frac{0.1\sqrt{11}}{\sqrt{0.6}}\right)\) \\
      & \(p_2\) &
        \(\left(\sqrt{0.6},-\sqrt{0.6},0,0\right)\) \\
      & \(p_3\) &
        \(\left(0,
        -\frac{0.5}{\sqrt{0.6}},
        0,
        -\frac{0.1\sqrt{11}}{\sqrt{0.6}}\right)\) \\
      & \(m\) & \((0,0,0,0)\) \\

    \multirow{4}{*}{\(P_5^\Box\)}
      & \(p_1\) &
        \(\left(
        -\sqrt{0.9},
        \frac{\sqrt{0.73}}{\sqrt{0.9}},
        0,
        \frac{0.1\sqrt{8}}{\sqrt{0.9}}
        \right)\) \\
      & \(p_2\) &
        \(\left(
        \sqrt{0.9},
        -\sqrt{0.9},
        0,
        0
        \right)\) \\
      & \(p_3\) &
        \(\left(
        0,
        -\frac{\sqrt{0.73}}{\sqrt{0.9}},
        0,
        -\frac{0.1\sqrt{8}}{\sqrt{0.9}}
        \right)\) \\
      & \(m\) & \((0,0,0,0)\) \\
    \addlinespace

       \multirow{4}{*}{\(P_6^\Box\)}
      & \(p_1\) &
        \(\left(
        -1,
        0.4,
        0,
        \sqrt{0.84}
        \right)\) \\
      & \(p_2\) &
        \(\left(
        1,
        -1,
        0,
        0
        \right)\) \\
      & \(p_3\) &
        \(\left(
        0,
        -0.4,
        0,
        -\sqrt{0.84}
        \right)\) \\
      & \(m\) & \((0,0,0,0)\)  \\
    \bottomrule
  \end{tabular}
   \caption{The considered kinematic points for the box integrals in figure~\ref{fig:topologies}.  Four-momenta are given in the convention
  \(p_i=(p_i^{(0)},p_i^{(1)},p_i^{(2)},p_i^{(3)})\) and \(m=(m_1,m_2,m_3,m_4)\) denotes the
  vector of internal masses. The configurations $P_1^\Box$, $P_2^\Box$, and $P_3^\Box$ yield convergent integrals for a trivial numerator. The corresponding numerical results are shown in table~\ref{tab:box-leading-results}. The integral associated with $P_4^\Box$ is divergent for a trivial choice of numerator. We use the numerator from eq.~\eqref{finitenumbox} that renders the integral finite and evaluate the resulting integral using our $hp$-quadrature algorithm. The results for the $P_4^\Box$ are shown in table~\ref{tab:box-numerator-result}.}
     \label{tab:box-points}
\end{table}

\begin{table}[h]
  \centering
  \small
  \renewcommand{\arraystretch}{1.15}
  \begin{tabular}{@{}ccl@{}}
    \toprule
    Configuration & Quantity & Value \\
    \midrule

    \multirow{5}{*}{\(P_1^{\smallpentagon}\)}
      & \(p_1\) & \((33.7452,45.7273,31.1525,-7.47943)\) \\
      & \(p_2\) & \((31.3644,-41.5073,46.479,2.04203)\) \\
      & \(p_3\) & \((4.59005,17.0701,32.654,41.9363)\) \\
      & \(p_4\) & \((29.5105,-28.2596,46.1733,-35.0892)\) \\
      & \(m\)   & \((5.01213,5.01213,5.01213,5.01213,5.01213)\) \\
    \addlinespace

    \multirow{5}{*}{\(P_2^{\smallpentagon}\)}
      & \(p_1\) & \((33.7648,45.4406,-10.6808,16.4193)\) \\
      & \(p_2\) & \((72.935,67.4917,-11.8149,-36.2846)\) \\
      & \(p_3\) & \((8.01673,-49.4011,-66.092,-0.11414)\) \\
      & \(p_4\) & \((-86.5419,-97.0123,68.1249,32.9488)\) \\
      & \(m\)   & \((98.427,28.8906,40.5144,75.4564,11.0833)\) \\

    \bottomrule
  \end{tabular}
     \caption{The considered kinematic points for the pentagon integral in figure~\ref{fig:topologies}.  Four-momenta are given in the convention
  \(p_i=(p_i^{(0)},p_i^{(1)},p_i^{(2)},p_i^{(3)})\) and \(m=(m_1,\ldots,m_5)\) denotes the
  vector of internal masses. Numerical results are shown in table~\ref{tab:pentagon-results}.}
       \label{tab:pentagon-points}
\end{table}

As a first example, we discuss the numerical evaluation of convergent one-loop integrals in $D=4$ dimensions in Euclidean kinematics. 
Hence, we simply need to evaluate the corresponding three-dimensional cLTD representation, cf. eq.~\eqref{LTDTriangleCausal} for the example of the triangle integral.
The results are listed in tables~\ref{tab:triangle-leading-results},~\ref{tab:box-leading-results} and~\ref{tab:pentagon-results} for the triangle, box and pentagon integrals, respectively. In order to gauge the efficiency of our $hp$-algorithm compared to {\tt Vegas}, we show the results obtained by both methods for (roughly) the same number of integrand evaluations. Overall, we find that for this number of evaluations, our $hp$-quadrature algorithm achieves the same precision as the analytic results implemented into {\tt LoopTools}, while {\tt Vegas} does not even reach half this precision. This illustrates that our $hp$-algorithm converges much faster towards the exact result than the MC approach.

\begin{table*}[h]
    \centering
    \small
    \setlength{\tabcolsep}{5pt}
    \renewcommand{\arraystretch}{1.15}

\begin{tabular}{@{}clll@{}}
    \toprule
    Configuration
    & \(hp\)-quadrature
    & \texttt{Vegas}
    & \texttt{LoopTools} \\
    \midrule

    \(P_1^\triangle\,[10^{-2}]\)
    & \(1.5420989685426680\)
    & \(1.54207 \pm 0.00001\)
    & \(1.5420989685426682\) \\
    &
    {\scriptsize\color{gray}\(N_{\mathrm{eval}}=65\,229\,725\)}
    &
    {\scriptsize\color{gray}
    \shortstack[l]{
    \(\varepsilon_{\mathrm{rel}}=7.406\times10^{-6}\)\\
    \(N_{\mathrm{eval}}=65\,016\,868\)}}
    & \\
    \addlinespace[3pt]

    \(P_2^\triangle\,[10^{-3}]\)
    & \(9.2489103207133699\)
    & \(9.24889 \pm 0.00003\)
    & \(9.2489103207133699\) \\
    &
    {\scriptsize\color{gray}\(N_{\mathrm{eval}}=66\,454\,634\)}
    &
    {\scriptsize\color{gray}
    \shortstack[l]{
    \(\varepsilon_{\mathrm{rel}}=2.820\times10^{-6}\)\\
    \(N_{\mathrm{eval}}=66\,451\,835\)}}
    & \\
    \addlinespace[3pt]

    \(P_3^\triangle\,[10^{-5}]\)
    & \(5.3636363075423607\)
    & \(5.36362 \pm 0.00004\)
    & \(5.3636363075423601\) \\
    &
    {\scriptsize\color{gray}\(N_{\mathrm{eval}}=66\,354\,781\)}
    &
    {\scriptsize\color{gray}
    \shortstack[l]{
    \(\varepsilon_{\mathrm{rel}}=7.578\times10^{-6}\)\\
    \(N_{\mathrm{eval}}=68\,722\,275\)}}
    & \\
    \addlinespace[3pt]

    \(P_4^\triangle\,[10^{-3}]\)
    & \(2.8206297586077672\)
    & \(2.82063 \pm 0.00001\)
    & \(2.8206297586077667\) \\
    &
    {\scriptsize\color{gray}\(N_{\mathrm{eval}}=64\,871\,471\)}
    &
    {\scriptsize\color{gray}
    \shortstack[l]{
    \(\varepsilon_{\mathrm{rel}}=4.161\times10^{-6}\)\\
    \(N_{\mathrm{eval}}=65\,097\,473\)}}
    & \\

    \bottomrule
\end{tabular}

    \caption{
        Numerical comparison for finite triangle integrals in the Euclidean
        region at the points specified in
        table~\ref{tab:triangle-points}. The gray numbers indicate the total
        number of integrand evaluations performed by the \(hp\)-quadrature and
        \texttt{Vegas}.
    The bracketed power
        of ten specifies the normalization of all numerical quantities in the
        corresponding row. For example, the entries in the first row have been
        multiplied by \(10^{5}\). The uncertainties following the
        \texttt{Vegas} values are the reported absolute error estimates in the
        same normalization.
    }
    \label{tab:triangle-leading-results}
\end{table*}

\begin{table*}[h]
    \centering
    \small
    \setlength{\tabcolsep}{5pt}
    \renewcommand{\arraystretch}{1.15}

\begin{tabular}{@{}cllll@{}}
    \toprule
    Configuration
    & \(hp\)-quadrature
    & \texttt{Vegas}
    & Analytic \\
    \midrule

    \(P_1^\triangle\,[10^{-2}]\),
    \(\mathcal{O}(\eps)\)
    & \(-5.7349778469781648\)
    & \(-5.734978 \pm 0.000004\)
    & \(-5.7349778469765868\) \\
    &
    {\scriptsize\color{gray}
        \(N_{\mathrm{eval}}=38\,457\,264\,704\)}
    &
    {\scriptsize\color{gray}
    \shortstack[l]{
        \(\varepsilon_{\mathrm{rel}}=6.534\times10^{-7}\)\\
        \(N_{\mathrm{eval}}=38\,475\,166\,371\)}}
    &
    & \\
    \addlinespace[3pt]

    \(P_1^\triangle\,[10^{-1}]\),
    \(\mathcal{O}(\eps^2)\)
    & \(1.2707120993052531\)
    & \(1.270713 \pm 0.000001\)
    & \(1.27071209930032053\) \\
    &
    {\scriptsize\color{gray}
        \(N_{\mathrm{eval}}=38\,457\,264\,704\)}
    &
    {\scriptsize\color{gray}
    \shortstack[l]{
        \(\varepsilon_{\mathrm{rel}}=1.024\times10^{-6}\)\\
        \(N_{\mathrm{eval}}=38\,500\,147\,054\)}}
    &
    & \\

    \bottomrule
\end{tabular}

    \caption{
        Numerical comparison for the first two higher-order coefficients
        in the \(\eps\)-expansion of the finite triangle integral in
        the Euclidean region at \(P_1^\triangle\), as specified in
        table~\ref{tab:triangle-points}.
        The bracketed power of ten specifies the normalization of all
        numerical quantities in the corresponding row; the displayed
        entries have been multiplied by \(10^{4}\).
        The uncertainties following the \texttt{Vegas} and
        \texttt{pySecDec} values are their reported absolute error
        estimates in the same normalization.
        The grey annotations give the \texttt{Vegas} relative error
        estimates and runtimes for the separate evaluation of each
        coefficient.
    }
    \label{tab:triangle-higher-epsilon-orders}
\end{table*}

\begin{table*}[h]
    \centering
    \small
    \setlength{\tabcolsep}{5pt}
    \renewcommand{\arraystretch}{1.15}
    \begin{tabular}{@{}clll@{}}
    \toprule
    Configuration
    & \(hp\)-quadrature
    & \texttt{Vegas}
    & \texttt{LoopTools} \\
    \midrule

    \(P_1^\Box\,[10^{-11}]\)
    & \(-3.4005008473604465\)
    & \(-3.4005 \pm 0.0001\)
    & \(-3.4005008473604458\) \\
    &
    {\scriptsize\color{gray}
        \(N_{\mathrm{eval}}=82\,215\,552\)}
    &
    {\scriptsize\color{gray}
    \shortstack[l]{
        \(\varepsilon_{\mathrm{rel}}=2.946\times10^{-5}\)\\
        \(N_{\mathrm{eval}}=82\,128\,251\)}}
    & \\
    \addlinespace[3pt]

    \(P_2^\Box\,[10^{-9}]\)
    & \(-2.1981438636707011\)
    & \(-2.19811 \pm 0.00002\)
    & \(-2.1981438636707011\) \\
    &
    {\scriptsize\color{gray}
        \(N_{\mathrm{eval}}=79\,581\,136\)}
    &
    {\scriptsize\color{gray}
    \shortstack[l]{
        \(\varepsilon_{\mathrm{rel}}=7.775\times10^{-6}\)\\
        \(N_{\mathrm{eval}}=79\,224\,393\)}}
    & \\
    \addlinespace[3pt]

    \(P_3^\Box\,[10^{0}]\)
    & \(-5.9148559407781782\)
    & \(-5.91482 \pm 0.00004\)
    & \(-5.9148559407781995\) \\
    &
    {\scriptsize\color{gray}
        \(N_{\mathrm{eval}}=78\,768\,061\)}
    &
    {\scriptsize\color{gray}
    \shortstack[l]{
        \(\varepsilon_{\mathrm{rel}}=5.996\times10^{-6}\)\\
        \(N_{\mathrm{eval}}=78\,043\,543\)}}
    & \\

    \bottomrule
\end{tabular}

    \caption{
        Numerical comparison for scalar box integrals in the Euclidean
        region at the points specified in
        table~\ref{tab:box-points}. The gray numbers indicate the total
        number of integrand evaluations performed by the \(hp\)-quadrature
        and \texttt{Vegas}.
        The bracketed power of ten specifies the normalization of all
        numerical quantities in the corresponding row. 
        The uncertainties following the \texttt{Vegas} values are the
        reported absolute error estimates in the same normalization.
    }
    \label{tab:box-leading-results}
\end{table*}

\begin{table*}[h]
    \centering
    \small
    \setlength{\tabcolsep}{5pt}
    \renewcommand{\arraystretch}{1.15}

    \begin{tabular}{@{}clll@{}}
    \toprule
    Configuration
    & \(hp\)-quadrature
    & \texttt{Vegas}
    & \texttt{LoopTools} \\
    \midrule

    \(P_1^{\smallpentagon}\,[10^{-11}]\)
    & \(1.9585377709960803\)
    & \(1.95853 \pm 0.00002\)
    & \(1.9585377709960810\) \\
    &
    {\scriptsize\color{gray}
        \(N_{\mathrm{eval}}=93\,600\,857\)}
    &
    {\scriptsize\color{gray}
    \shortstack[l]{
        \(\varepsilon_{\mathrm{rel}}=7.734\times10^{-6}\)\\
        \(N_{\mathrm{eval}}=93\,158\,247\)}}
    & \\
    \addlinespace[3pt]

    \(P_2^{\smallpentagon}\,[10^{-12}]\)
    & \(2.3427532470686744\)
    & \(2.342740 \pm 0.000009\)
    & \(2.3427532470686740\) \\
    &
    {\scriptsize\color{gray}
        \(N_{\mathrm{eval}}=96\,977\,644\)}
    &
    {\scriptsize\color{gray}
    \shortstack[l]{
        \(\varepsilon_{\mathrm{rel}}=3.853\times10^{-6}\)\\
        \(N_{\mathrm{eval}}=95\,838\,495\)}}
    & \\

    \bottomrule
\end{tabular}

    \caption{
        Numerical comparison for pentagon integrals in the Euclidean
        region at the points specified in
        table~\ref{tab:pentagon-points}. The gray numbers indicate the total
        number of integrand evaluations performed by the \(hp\)-quadrature
        and \texttt{Vegas}. For a meaningful comparison, both methods are
        run with approximately the same number of integrand evaluations.
        The bracketed power of ten specifies the normalization of all
        numerical quantities in the corresponding row.
        The uncertainties following the \texttt{Vegas} values are the
        reported absolute error estimates in the same normalization.
    }
    \label{tab:pentagon-results}
\end{table*}

We can also apply our algorithm to compute the higher orders in the $\eps$-expansion of finite one-loop integrals. We focus on the one-loop triangle for simplicity, though there are no obstacles to extend these conclusions to boxes and pentagons.
We start from eq.~\eqref{higherordersgen}, which in our case we may cast in the form: 

\begin{align}\label{trihighereps}
I^{D_{\eps}}_\triangle=&\frac{2^{-\eps}}{\Gamma(1-\eps)}I_\triangle^{D}(0)\\
&+\frac{1}{\Gamma(-\epsilon)}\int_0^1 \rd t\, \left(\left(\frac{(1-t)^2}{(2-t)t}\right)^\epsilon\left(\frac{1}{t} I_\triangle^{D}(t)+\frac{2}{1-t}I_\triangle^{D}(t)\right)-2^{-\epsilon} I_\triangle^{D}(0)\right)\notag\\
&-\frac{1}{\Gamma(-\eps)}\int_0^1 \rd t\, \frac{1}{2-t}\left(\frac{(1-t)^2}{(2-t)t}\right)^\eps I_\triangle^{D}(t)\notag\,.
\end{align}
Each of the integrals in the right-hand side of eq.~\eqref{trihighereps} is convergent for $\eps=0$, and can be expanded into a series in $\eps$ under the integral sign. We then insert the cLTD representation of the one-loop triangle in $D=4$ dimensions (cf.~eq.~\eqref{triangleLTD}), with masses shifted by $\eta$, where $\eta=\frac{1}{(t-1)^2}-1$, and we apply the $hp$-quadrature algorithm to the combined four-dimensional integral over $t$ and the spatial part of the loop momentum coming from the cLTD representation in eq.~\eqref{triangleLTD}.
Indeed, since all proofs in section~\ref{sec:hp-quadrature} are valid for an arbitrary number of dimensions $d$, we can extend our $hp$-quadrature to fourfold integrals like in eq.~\eqref{trihighereps}, and we need to include the singularity at $t=0$ into our set of singular points. We compare our results for the first two higher orders in $\epsilon$ to numerical results obtained by {\tt Vegas} and {\tt PySecDec}~\cite{Borowka:2017idc}, as well as to an analytic result in terms of polylogarithms obtained via direct integration techniques (cf.,~e.g., ref.~\cite{Chavez:2012kn}), for which we use \texttt{PolyLogTools}~\cite{Duhr:2019tlz}. The results can be found in table~\ref{tab:triangle-higher-epsilon-orders}. 

Until now, our choice of examples was restricted to convergent one-loop integrals with unit numerators. However, not all one-loop integrals of this type are finite, and there are one-loop triangle, box and pentagon integrals that are divergent in four dimensions. We now illustrate on a the example of the one-loop box integral from figure figure~\ref{fig:topologies} with two external masses and $p_1^2=p_2^2=0$ how we can apply the cLTD approach with the construction of numerators that render the integrals finite so that they can be evaluated using our $hp$-quadrature algorithm.
We use the method of ref.~\cite{Gambuti:2023eqh} to construct numerators that render this integral
UV and IR finite. They have at most overall degree five and at most degree three in the individual loop momenta. 
A possible such numerator reads:
\begin{align}\label{finitenumbox}
\mathcal{N}_\Box^{\scriptscriptstyle(\text{2m})}&=\frac{1}{s_{12}^2 \left(M_3^4-2 M_3^2 s_{13}+s_{13} (s_{12}+s_{13})\right)^3}\Big\{[k\cdot p_1 (M_3^2-s_{12}-s_{13})\notag\\
&+k\cdot p_2 (s_{13}-M_3^2)-s_{12} k\cdot p_3] \Big[k\cdot p_1 (M_3^2-s_{13}) (M_3^2-s_{12}-s_{13})\\
&+k\cdot p_2 \left(M_3^4-2 M_3^2 s_{13}+2 s_{12} s_{13}+s_{13}^2\right)+s_{12} k\cdot p_3 (s_{13}-M_3^2)\Big]\notag \\
&\times \Big[k\cdot p_1 (-M_3^2+s_{12}+s_{13})^2+k\cdot p_2 \left(M_3^4+M_3^2 (s_{12}-2 s_{13})+s_{13} (s_{12}+s_{13})\right)\notag\\
&+s_{12} k\cdot p_3 (-M_3^2+s_{12}+s_{13})\Big]\Big\}\notag\,,
\end{align}
with $M_3^2=p_3^2$. It is easy to check that the box integral with this numerator is finite using integration-by-parts (IBP) identities~\cite{Chetyrkin:1981qh,Tkachov:1981wb} to express this integral in terms of the standard (divergent) basis of integrals with unit numerators. 
We can then derive an cLTD representation for the integral with this numerator. Since the integral is by construction finite, we can evaluate it numerically using our $hp$-quadrature algorithm.
The results are shown in in table~\ref{tab:box-numerator-result}. We observe roughly the same pattern as for the convergent integrals with unit numerators. Our $hp$-quadrature algorithm converges faster to the exact result than the MC implementation.
\begin{table*}[h]
    \centering
    \small
    \setlength{\tabcolsep}{5pt}
    \renewcommand{\arraystretch}{1.15}
\begin{tabular}{@{}cllll@{}}
    \toprule
    Configuration
    & \(hp\)-quadrature
    & \texttt{Vegas}
    & Analytic \\
    \midrule
    \(P_4^\Box\,[10^{-1}]\)
    & \(3.4669191413352562\)
    & \(3.4672 \pm 0.0004\)
    & \(3.4669191413352407\) \\
    &
    {\scriptsize\color{gray}
    \shortstack[l]{
        \(N_{\mathrm{eval}}=2\,883\,120\,744\)}}
    &
    {\scriptsize\color{gray}
    \shortstack[l]{
        \(\varepsilon_{\mathrm{rel}}=1.287\times10^{-4}\)\\
        \(N_{\mathrm{eval}}=2\,867\,196\,421\)}}
    &
    &
    \\
    \addlinespace[3pt]
    \(P_5^\Box\,[10^{-1}]\)
    & \(8.1029284279864500\)
    & \(8.1030 \pm 0.0002\)
    & \(8.1029284279864533\) \\
    &
    {\scriptsize\color{gray}
    \shortstack[l]{
        \(N_{\mathrm{eval}}=3\,057\,407\,913\)}}
    &
    {\scriptsize\color{gray}
    \shortstack[l]{
        \(\varepsilon_{\mathrm{rel}}=2.311\times10^{-5}\)\\
        \(N_{\mathrm{eval}}=3\,059\,900\,041\)}}
    &
    &
    \\
    \addlinespace[3pt]
    \(P_6^\Box\,[10^{-2}]\)
    & \(-1.4532452983434587\)
    & \(-1.4535 \pm 0.0009\)
    & \(-1.4532452983434252\) \\
    &
    {\scriptsize\color{gray}
    \shortstack[l]{
        \(N_{\mathrm{eval}}=3\,969\,467\,212\)}}
    &
    {\scriptsize\color{gray}
    \shortstack[l]{
        \(\varepsilon_{\mathrm{rel}}=5.851\times10^{-4}\)\\
        \(N_{\mathrm{eval}}=3\,968\,742\,980\)}}
    &
    &
    \\
    \bottomrule
\end{tabular}
    \caption{
        Numerical comparison for box integrals with a numerator in the
        Euclidean region at the points specified in
        table~\ref{tab:box-points}. The gray numbers indicate the total
        number of integrand evaluations performed by the \(hp\)-quadrature
        and \texttt{Vegas}.
        The bracketed power of ten specifies the normalization of all
        numerical quantities in the corresponding row. The uncertainties
        following the \texttt{Vegas} values are
        their reported absolute error estimates in the same normalization.  }
    \label{tab:box-numerator-result}
\end{table*}

 \subsection{Towards $hp$-quadrature in the physical region}
\label{sec.NonEucl}

So far, we have only considered one-loop integrals in Euclidean kinematics, for which  $E$-surfaces do not intersect the integration domain.
Outside of this region, this may no longer be the case.
While we leave a more complete treatment of threshold singularities within our approach to future work~\cite{hp-quad-imag}, in this section we present a first proof-of-concept that it is possible to obtain fast and stable numerical results with our $hp$-quadrature algorithm also in the case of threshold singularities.

 To obtain the correct imaginary parts, we use \textit{threshold subtraction} (cf., e.g.,~refs.~\cite{Kermanschah:2021wbk, Kermanschah:2025wlo, Kermanschah:2024utt,Kilian:2009wy,Aguilera-Verdugo:2020set}). If there is only one threshold crossed, the real part corresponds to the principal value of the integral and the imaginary part can be obtained by inserting a $\delta$-function. 
 Due to the compactness of the $E$-surfaces, the  principle value part is  well suited to be integrated in the $hp$-quadrature formalism and we illustrate this approach for a simple example.

\begin{table}[!t]
    \centering
    \small
    \renewcommand{\arraystretch}{1.15}

    \begin{tabular}{@{}ccl@{}}
        \toprule
        Configuration & Quantity & Value \\
        \midrule

        \multirow{3}{*}{\(P_5^\triangle\)}
          & \(p_1\) & \((44.3894,0,0,0)\) \\
          & \(p_2\) & \((0.62982,-17.8442,17.8442,23.6744)\) \\
          & \(m\)   & \((7.89824,70.8982,7.89824)\) \\
        \addlinespace

        \multirow{3}{*}{\(P_6^\triangle\)}
          & \(p_1\) & \((70,0,0,0)\) \\
          & \(p_2\) & \((1,0.2,100,250)\) \\
          & \(m\)   & \((30,40,30)\) \\
        \addlinespace

        \multirow{3}{*}{\(P_7^\triangle\)}
          & \(p_1\) & \((42.7,0,0,0)\) \\
          & \(p_2\) & \((7.79,7000.97,320.2,320.1)\) \\
          & \(m\)   & \((20.79,52.42,20.79)\) \\

        \bottomrule
    \end{tabular}

    \caption{
        Kinematic points for triangle integrals above the threshold used in
        the numerical evaluations.
        The corresponding numerical results are given in
        table~\ref{tab:triangle-threshold-results}.
        Four-momenta are specified using the convention
        \(p_i=(p_i^{(0)},p_i^{(1)},p_i^{(2)},p_i^{(3)})\), while
        \(m=(m_1,m_2,m_3)\) denotes the vector of internal masses.
    }
    \label{tab:triangle-points-aboveThreshold}
\end{table}
To this effect, we evaluate the triangle diagram in eq.~\eqref{introtriang} at a  kinematic point outside the Euclidean region defined in table~\ref{tab:triangle-points-aboveThreshold}.
This kinematic point is chosen such that the zero locus of $E:=E_{31}={E_{1}^{\scriptscriptstyle(-,0)}+E_3}$ is contained in $\mathbb{R}^3$.
We first write the cLTD integrand in terms of  spherical coordinates $(r,\theta,\phi)$. For this particular kinematic point, expressed in spherical coordinates, $E$ only depends on $r$,
\begin{align}\label{Esurf}
2 \sqrt{m_3^2+r^2}-p_1^{(0)}=E\,.
\end{align}
We solve eq.~\eqref{Esurf} for $r$ as
\begin{align}\label{rrrep}
r=\pm\frac{1}{2}\sqrt{-4m_3^2+\left(E+p_1^{(0)}\right)^2}\,,
\end{align}
and we change variables from $r$ to $E$.
For our particular choice of kinematics, we have $(p_1^{\scriptscriptstyle(0)})^2>4 m_3^2$,such that the square root in eq.~\eqref{rrrep} is well defined in the entire integration domain. Further, since the radial integration domain is restricted to $r>0$, only the positive solution in eq.~\eqref{rrrep} is relevant.
After this change of variables, the integral in eq.~\eqref{trianglesurfaces} can be cast in the form $I_{\triangle} = I_{\triangle\text{, sing}}+I_{\triangle\text{, reg}}$, where $I_{\triangle\text{, reg}}$ is integrable in four dimensions. The singular terms, which require careful treatment, are 
\begin{align}
I_{\triangle\text{, sing}}&=\frac{1}{4 \pi}\int_{2m_3-p_1^{(0)}}^\infty \rd E\, \mathcal{J}_E \int_{0}^\pi \rd \theta \int_{0}^{2 \pi} \rd \phi\, \sin{\theta}\,\frac{f(E,\theta,\phi)}{E}\notag\,,
\end{align}
with the Jacobian 
\begin{equation}
    \mathcal{J}_E= \frac{1}{2} r \sqrt{m_3^2+r^2}=\frac{\sqrt{\left(E+p_1^{(0)}\right)^2}}{8}\sqrt{-4m_3^2+\left(E+p_1^{(0)}\right)^2}\,,
\end{equation}
and 
\begin{align}
f(E,\theta,\phi)=\frac{1}{E_1 E_2 E_3 }\Bigg(\frac{1}{E_{32}}+\frac{1}{E_{21}}\Bigg)\,.
\end{align}
According to Sokhotski–Plemelj theorem~\cite{Plemelj1908,Sokhotski1873}, which states  that \[
\frac{1}{x \pm i0}
=
\operatorname{PV}\!\left(\frac{1}{x}\right)
\mp i\pi\,\delta(x)\,,
\]
the integral of the singular contribution is given by
\begin{align}
I_{\triangle\text{, sing}}&=\operatorname{PV}\left[I_{\triangle\text{, sing}}\right]+i \pi \frac{1}{4 \pi}\int \rd \theta \,\rd E\,\rd \phi\, \mathcal{J}_E \,\sin{\theta}\,\delta\!\left(E\right)\,f(E,\theta,\phi)\,.
\end{align} 
Here, $\operatorname{PV}$ denotes the principal value.
The real and imaginary parts of the triangle integral are then given by 
\begin{equation}\begin{split}
\label{TriImPart}
\mathrm{Re}[I_{\triangle}]&=I_{\triangle\text{, reg}}+\operatorname{PV}\left[ I_{\triangle\text{, sing}}\right]=\left(I_\triangle-I_{\triangle\text{, sing}}\right)+\operatorname{PV}\left[ I_{\triangle\text{, sing}}\right]\,,\\
\text{Im}\left[I_\triangle\right]
&=\frac{i \pi }{4 \pi}  \int_0^\pi \rd \theta \int_0^{2\pi}\rd \phi  \,\sin{\theta}\,\mathcal{J}_E \,f(0,\theta,\phi)\,.
\end{split}\end{equation}
 Both the real and imaginary parts can now be integrated numerically. For the real part, we employ our $hp$-quadrature algorithm, while the imaginary part is only a two-dimensional integral and can efficiently be integrated using quadrature (without refinement). The result is given in table~\ref{tab:triangle-threshold-results}. Just like for the examples in Euclidean kinematics, we observe a much faster convergence in the number of function evaluations of our $hp$-quadrature  algorithm than for MC integration.
 \begin{table*}[h]
    \centering
    \small
    \setlength{\tabcolsep}{4pt}
    \renewcommand{\arraystretch}{1.15}
\begin{tabular}{@{}cllll@{}}
    \toprule
    Configuration
    & Component
    & \(hp\)-quadrature
    & \texttt{Vegas}
    & \texttt{LoopTools} \\
    \midrule

    \(P_5^\triangle\)
    & \(\operatorname{Re}[10^{-4}]\)
    & \(3.9847997696028542\)
    & \(3.984799 \pm 0.000001\)
    & \(3.984799769602855\) \\
    &  &         
    {\scriptsize\color{gray}
        \shortstack[l]{
            \(N_{\mathrm{eval}}={164\,347\,506}\)}} &
     \scriptsize\color{gray}
    \shortstack[l]{
    \(\varepsilon_{\mathrm{rel}}=3.076\times10^{-7}\)\\
    \(N_{\mathrm{eval}}=1\,638\,897\,491\)}
    & \\[2pt]

    &
    \(\operatorname{Im}[10^{-4}]\)
    & \(4.8774154335541110\)
    & \(4.87741554 \pm 0.00000008\)
    & \(4.8774154335541111\) \\
    & & &
  \scriptsize\color{gray}
    \shortstack[l]{
    \(\varepsilon_{\mathrm{rel}}=1.621\times10^{-8}\)\\
    \(N_{\mathrm{eval}}=1\,573\,458\,369\)}
    & \\
    \addlinespace[4pt]

    \(P_6^\triangle\)
    & \(\operatorname{Re}[10^{-5}]\)
    & \(7.8949736287057560\)
    & \(7.894974 \pm 0.000002\)
    & \(7.8949736287057561\) \\
    &  &         
    {\scriptsize\color{gray}
        \shortstack[l]{
            \(N_{\mathrm{eval}}={183\,204\,250}\)}} &
    \scriptsize\color{gray}
    \shortstack[l]{
    \(\varepsilon_{\mathrm{rel}}=2.936\times10^{-7}\)\\
    \(N_{\mathrm{eval}}=1\,870\,215\,721\)}
    & \\[2pt]

    &
    \(\operatorname{Im}[10^{-5}]\)
    & \(2.2258948981264372\)
    & \(2.22589492 \pm 0.00000003\)
    & \(2.2258948981264375\) \\
    & & &
  \scriptsize\color{gray}
    \shortstack[l]{
    \(\varepsilon_{\mathrm{rel}}=1.478\times10^{-8}\)\\
    \(N_{\mathrm{eval}}=1\,754\,481\,990\)}
    & \\
    \addlinespace[4pt]

    \(P_7^\triangle\)
    & \(\operatorname{Re}[10^{-7}]\)
    &   \(2.7502744387947264\)
    & \(2.750275 \pm 0.000001\)
    & \(2.7502744387947350\) \\
    &  &         
    {\scriptsize\color{gray}
        \shortstack[l]{
            \(N_{\mathrm{eval}}={196\,056\,782}\)}} &
      \scriptsize\color{gray}
    \shortstack[l]{
    \(\varepsilon_{\mathrm{rel}}=3.763\times10^{-7}\)\\
    \(N_{\mathrm{eval}}=1\,979\,057\,741\)}
    & \\[2pt]

    &
    \(\operatorname{Im}[10^{-8}]\)
    &   \(1.4522706831607231\)
    & \(1.45227068 \pm 0.00000002\)
    & \(1.452270683160722\) \\
    & & &
  \scriptsize\color{gray}
    \shortstack[l]{
    \(\varepsilon_{\mathrm{rel}}=1.427\times10^{-7}\)\\
    \(N_{\mathrm{eval}}=1\,853\,446\,284\)}
    & \\

    \bottomrule
\end{tabular}

    \caption{
        Numerical comparison of the real and imaginary parts above the
        threshold at the points \(P_5^\triangle\), \(P_6^\triangle\), and
        \(P_7^\triangle\), specified in
        table~\ref{tab:triangle-points-aboveThreshold}.
        The \(hp\)-quadrature and \texttt{Vegas} results are compared with
        the corresponding analytic values evaluated using
        \texttt{LoopTools} in quadruple precision.
        The bracketed power of ten specifies the normalization of all
        numerical quantities in the corresponding row; for example, the
        entries in the first row have been multiplied by \(10^6\).
    }
    \label{tab:triangle-threshold-results}
\end{table*}

\section{Conclusions and Outlook}
\label{sec.conclusion}
In this paper we have introduced a numerical algorithm to evaluate one-loop integrals in dimensional regularisation to any desired order in the $\eps$-expansion. In order to achieve this, we have combined various advances from both mathematics and physics. On the physics side, we use the cLTD representation of one-loop integrals, and we render divergent integrals finite by introducing appropriate numerator factors. Higher orders in $\eps$ are obtained by integrating over an auxiliary propagator mass. In this way, we obtain at every order in $\eps$ an (at most) fourfold convergent integrals representation whose singularities are all located on a compact set. From the mathematics side, such a representation is ideally suited for numerical integration using $hp$-quadrature. We have provided a rigorous proof that $hp$-quadrature is applicable to our integrals, and that our algorithm converges with a stretched exponential convergence in the number of integrand evaluations. 

We have implemented our algorithm into a numerical code, and we have benchmarked it against Monte Carlo integration as well as existing analytic results for triangle, box and pentagon integrals evaluated at selected Euclidean kinematic points. For all cases, we observe that our algorithm converges much faster than Monte Carlo integration.

For the future, we plan to extend our algorithm to include also physical kinematic points. This requires us to subtract the threshold singularities that develop in the physical scattering region. We have included a proof of concept evaluation of a triangle integral in physical kinematics. While this preliminary study shows that our algorithm can be extended to physical kinematics, we leave a complete implementation for future work. Once this is completed, we have a numerical algorithm that evaluates  one-loop integrals with up to five external legs numerically to any desired order in $\eps$, which is an important ingredient for computations beyond one loop.

\section*{Acknowledgments}
\label{sec.acknowledgments}
We are grateful to Zeno Capatti and Aniruddha Venkata for helpful discussions.
This work is supported in part by the the
Deutsche Forschungsgemeinschaft (DFG, German Research Foundation) under Germany’s
Excellence Strategy – EXC 3107 – Project-ID 53376636 and EXC-2047/2 – Project-ID 390685813, as well as the CRC 1639 “NuMeriQS” – Project-ID 511713970.

\begin{appendix}
\section{Integration on the residual domain}\label{app:resdom}
In this appendix, we discuss the integration of the residual unbounded $D_R$ domain without any singularities, see \cref{eq:residualsplitting} for its definition. For the numerical integration on $D_R$ we need to transform $D_R$ into a bounded domain. We proceed by splitting $D_R$ into two domains
\[
D_R := D_G \cup D_U\,,
\qquad
\text{where}
\qquad
D_G := B_{2b\sqrt{d}} \backslash D_T, \quad D_U := \mathbb{R}^d \backslash B_{2b\sqrt{d}}\,,
\]
and
\[
B_r := \left\{ \mathbf{x} \in \mathbb{R}^d : \| \mathbf{x}\|_2 \leq r \right\}.
\]

Let $T_S\colon[0,1]^d\to B_1$ be the coordinate transform from spherical to Cartesian coordinates on $B_1$. Then the integral over $D_U$ can be reformulated to an integral over $[0,1]^d$ by 
\begin{equation}\label{eqn:transform_unbound_bound}
\begin{aligned}
\int_{D_U} f(\mathbf{x}) \,\rd \mathbf{x} &= \int_{2 b \sqrt{d}}^\infty \int_{[0,1]^{d-1}} f(T_S (r, \boldsymbol{\varphi})) |\operatorname{det} J_{T_S}(r, \boldsymbol{\varphi})| \,\rd\boldsymbol{\varphi} \,\rd r \\ &= \int_{[0,1]^d} f\left(T_S \left(\frac{2b \sqrt{d}}{1-\hat{r}}, \boldsymbol{\varphi} \right)\right) \frac{2b \sqrt{d}}{(1-\hat{r})^2} \left|\operatorname{det} J_{T_S} \left(\frac{2b\sqrt{d}}{(1-\hat{r})^2}, \boldsymbol{\varphi} \right) \right| \,\rd \boldsymbol{\varphi} \,\rd \hat{r}\,.
\end{aligned}
\end{equation}
This transformation introduces a singularity at $\hat{r}=1$. Numerically, this can be dealt with by applying $hp$-quadrature in the coordinate direction of $r$ and Gauss-Legendre quadrature in the other coordinate directions.

Integration of $D_G$ is slightly more difficult as we need to integrate an area between a hypercube and a $d$-sphere. To this end, observe that
\[
D_G= \left\{c\mathbf{x} : \mathbf{x} \in \partial D_T, c \in [1,2b \sqrt{d} \|\mathbf{x}\|^{-1}]  \right\},
\]
and that $D_G$ can be further subdivided into $D_G=\cup_{i=1}^{2d} V_i$, where $V_i$ corresponds to a projection from one side of hypercube $D_T$. These $V_i$ are trivially parametrized over $[0,1]^d$, to which tensor product Gauss-Legendre quadrature can be applied.

\section{Choice of $b$}\label{app:choiceofb}
In this appendix we discuss the choice of the constant $b$ from section~\ref{sec.hp}.
Technically, any $b$ such that $D_T$ from eq.~\eqref{eq:defb} contains $S$ would be sufficient to obtain stretched-exponential quadrature convergence on the residual domain $D_R$, but one natural characterization exists, which depends on the unbounded to bounded domain transform in eq.~(\ref{eqn:transform_unbound_bound}). As shown, if we wish to map the norm of each point in $D_U$ to a reference interval [0,1], the mapping is defined as: $2b\sqrt{d}/(1-r)$. If we take the furthest point in $P$ from the origin, in the resulting domain, the singularity is mapped to
 \[r_p := 1 - \frac{2b\sqrt{d}}{\max_{p \in P}\|p\|_2}\,.\]
 Following the quadrature error scaling expressed, it is desirable to $r_p$ to satisfy $r_p<-\sqrt{d}$. Because we are also integrating the domain $D_G$, we will double this distance. In addition we want to ensure that $b\geq1$ as a condition from lemma~\ref{lemma:minimum_refinment_levels}:
\[
b := \max\left(1, \max_{p\in P}\|p\|_2 \frac{1+\sqrt{d}}{\sqrt{d}} \right).
\]
\end{appendix}
\bibliographystyle{JHEP}
\bibliography{refs.bib}

\end{document}